\documentclass{aa}
\usepackage{graphicx}

\begin{document}

%\thesaurus{11.05.2; 11.12.2}

\title{The Luminosity Function Of Field Galaxies And Its Evolution 
Since $z=1$ }

\author{J.W. Fried, B. von Kuhlmann, K. Meisenheimer, H.-W. Rix, 
C. Wolf, H.H. Hippelein, M. K\"ummel, S. Phleps, H.J. R\"oser, 
I. Thierring, C. Maier}

\offprints{J.W. Fried}

\institute{Max-Planck-Institut f\"ur Astronomie, K\"onigstuhl 17, D-69117 Heidelberg, Germany\\
\email{fried@mpia-hd.mpg.de}}

\date{Received date; accepted date}

\abstract{
We present the B-band luminosity function and comoving 
space and luminosity densities for a sample of 
2779 I-band selected field galaxies based 
on multi-color data from the CADIS survey. The sample is complete down 
to $I_{815} = 22$ without correction and with completeness correction 
extends to $I_{815}=23.0$. By means of a new multi-color 
analysis the objects are classified according to their spectral energy distributions 
(SEDs) and their redshifts are determined  with typical errors of $\delta z \le 0.03$. 
We have split our sample into four redshift bins between $z=0.1$ and $z=1.04$ 
and into three SED bins E-Sa,Sa-Sc and starbursting (emission line) galaxies.
The evolution of the luminosity function is clearly differential with SED.  
The  normalization $\phi^*$ of luminosity function for the E-Sa galaxies 
decreases towards higher redshift, and  we find evidence that the comoving galaxy 
space density decreases with redshift as well. In contrast, 
we find  $\phi^*$ and the comoving space density increasing with redshift 
for the Sa-Sc galaxies. For the starburst galaxies we find a steepening of 
the luminosity function at the faint end and their comoving space density 
increases with redshift.
}
\authorrunning{Fried et al.}
\titlerunning{Evolution of the field galaxy luminosity function}

\maketitle

\section{Introduction}

The luminosity function, specifying the density of galaxies within a given 
comoving volume as a function of type and magnitude, is  a basic ingredient 
in describing the galaxy population. A determination of the luminosity function 
at different redshifts  describes in a global way the evolution of the 
galaxy population with cosmic look back time. Luminosity function 
data, combined with other data such as number counts and 
direct imaging data, form the observational basis for probing  
the statistical evolution of galaxies.

The present epoch  ($z < 0.2$) luminosity function is important 
as reference point of the evolution; there has been much controversy 
about its shape and normalization, but now it appears well established 
( \cite{zucca}) that the faint end $M_B \ge -16$ of the luminosity function 
is dominated by galaxies of later morphology, later spectral type, bluer 
color and stronger line emission.

To map galaxy  evolution over the last half of the universe's age, 
three surveys with samples of several hundred galaxies have been 
published in the last years: 
the CFRS survey (\cite{lilly95}) with 591 I-band selected galaxies with 
measured redshifts extending out to $z \approx 1$, 
the autofib survey (\cite{ell96}) with $\ge 1700$ galaxies 
with redshifts $z < 0.75$,
and the CNOC2 survey (\cite{cnoc2}) with over 2000 R-band selected 
galaxies with redshifts $z \le 0.55$. \cite{lilly95} found no evolution
of the luminosity function for galaxies earlier than Hubble type Sb but a 
steepening of the faint-end luminosity function with increasing redshift 
for later galaxy types, i.e. the galaxy population is richer in 
dwarfs with increasing redshift. Such a steepening of the luminosity 
function has also been found by \cite{ell96}. 
In contrast, the CNOC2 survey claimed to find  luminosity evolution  
for early type galaxy luminosity function and density evolution for 
the late type galaxy luminosity function, which is not in agreement with the CFRS results. 
\cite{Kauff96} found strong evolution in the population of the 
early type galaxies by re-analyzing the CFRS data, in the sense that at $z=1$ two-thirds 
of the nearby early type galaxy population have dropped out of the sample. 
However, \cite{tot98} did not confirm this conclusion, but found that the 
data are consistent with passive evolution. 

In this paper we present an independent determination of 
the luminosity function based on a new, large sample of galaxies. 
The CFRS, autofib and CNOC2 surveys are based on direct imaging data 
from which objects are separated into stars and galaxies based on 
morphology; then redshifts are measured for galaxies down to a 
certain limiting magnitude in a given band. As many faint galaxies 
are barely resolved, size selection 
effects, which can not  be easily understood or modelled, enter in 
the determination of the luminosity function. The approach  we take  
here is completely different. We use multi-color data and determine 
galaxy type and redshift from comparison to a library of color indices. 
This automatic process can be simulated and therefore we can reliably 
determine the completeness of our data. A further advantage of our 12-color data 
is that we do not have to apply k-corrections by fits of polynomials or 
the like; rather we interpolate measured data. The disadvantage of our 
multi-color method are redshift errors which are larger than spectroscopic ones, 
but we can show that they do not affect the luminosity function.

\section{The Data Sample} \label{sample}

Since CADIS will be described in detail elsewhere 
(Meisenheimer et al., in preparation), we only briefly summarize 
the most pertinent information necessary for this paper. 
Once finished, CADIS will cover seven $100 \sq \arcmin$ fields 
in different regions of the sky which avoid obvious foreground 
stars and for which the IRAS cirrus at $100 \mu m$ is $ \le 2 MJy/sr$. 
Our present analysis is based on three widely separated fields 
(1h, 9h and 16h field, see table \ref{fields}).

\begin{table}[here] \label{fields}
\caption[]{The centers of the fields}
\begin{flushleft}
\begin{tabular}{rrr} \hline
field  & $ \alpha_{2000} $  & $\delta_{2000}$ \\ \hline
1 h  & $1^h 47^m 33 \fs 3$  & $2 \degr 19 \arcmin 55 \arcsec$     \\
9 h  & $9^h 13^m 46 \fs 5 $ & $46 \degr 14 \arcmin 20 \arcsec    $\\
16 h& $16^h 24^m 32 \fs 3$  & $55 \degr 44\arcmin  32 \arcsec  $     \\ \hline
\end{tabular}
\end{flushleft}
\end{table}

CADIS is a multi-color survey, using three broad band 
(B, R, J or K') and up to 13 medium band ($\delta \lambda = 250$ to $500  \AA $) 
filters with central wavelengths from $3900$ to $9150  \AA $. Although 
these filters are tailored for detection of high redshift emission line objects, 
the resulting multi-color photometry allows excellent classification of the 
objects according to their SEDs and determination of their redshifts.

Total integration  times in a narrow band filter are on the order 
of 20 ksec on the 2.2m telescope, split into several 
dithered exposures; these long exposure times result in 
$10 \sigma$ detection  limits of $m \ge 23.0$ in {\em all} filters.

For object detection we use the SExtractor software package (\cite{ber96}) 
on the coadded frames for each filter. Photometry of 
the detected objects is done with our own software package MPIAphot. 
This photometry package optimizes the signal-to-noise ratio by deriving 
the fluxes above local background with a weighted sum and taking 
seeing variations into account (\cite{romei91}, \cite{thom96}). 
Since the photometry is an aperture photometry with  apertures  1.3 times 
seeing measured on the individual frame, the magnitudes of large objects must be 
corrected for aperture losses. These corrections, which are dependent on the morphology 
of the galaxy,  were derived from photometry on 
simulated images where the true magnitudes were known.
Because the photometry is performed on  individual rather than on coadded frames, 
realistic estimates of the photometric errors can be derived straightforwardly. 
Conversion of fluxes to magnitudes is based on  observations of spectrophotometric 
standard stars and synthetic photometry, i.e. integration of the instrumental response 
function over the standard star spectrum. From the loci of the measured 
stellar main sequence we estimate that the  systematic errors in absolute photometry 
are less than 0.03 mag in all filters. The magnitudes we use are Vega magnitudes.  
For each object our photometric catalogue thus contains (among other data) 
10 to 13 magnitude entries, their errors and 9 to 12 color indices.

The color indices in the object catalogues are used to classify 
the objects into stars, galaxies and QSOs by comparing them to a 
color index library derived from the SEDs of stars, galaxies and QSOs. 
Note that we do not apply any morphological star/galaxy separation 
or use other criteria, the classification is entirely spectroscopic. 
The stellar library is taken from \cite{pickles}. The QSO spectral library is based 
on the high signal-to-noise template spectrum \cite{fran}, but also 
includes different continuum slopes and line-widths; the final 
QSO spectral library contains templates with redshifts up to $z \le 6$. 
The spectral library for galaxies is derived 
from the mean averaged  spectra of \cite{kinn}. From these, a 
grid of redshifted template spectra was formed covering redshifts 
from $z = 0$ to $z = 2$ in steps of $ \delta z = 0.01 $ 
and 100 different intrinsic SEDs,  from old populations to starbursts. 
Our final library of color indices 
contains  entries for 131 star-, 45150 QSO- and 20100 galaxy templates.  
Using the minimum error variance estimator, each object is assigned  a type 
(star, QSO, galaxy), a redshift (if it is not classified as star) 
and an SED. The formal errors in this process depend on magnitude 
and type of the object and  are on the order of $\delta z = 0.03$ and 
$\delta SED =2$ (of 100). The dispersion in the estimated values 
does not increase with redshift for $z \le 1.04$. Full details are 
given in \cite{wolf}.

Down to the magnitude limit $m_{I815} \ge 23.0$ in a medium band filter 
centered at $\lambda = 815nm$, 4303 objects were detected, 
835 of these were classified as stars, 69 as QSOs and 3399 as galaxies, 2779 
of which are within the redshift range $z=0.1$ and $z=1.04$. 
For 257 objects (i.e. 6 \%) neither redshift nor SED could be estimated.

\begin{figure}[h]
 \resizebox{\hsize}{!}{\includegraphics[clip]{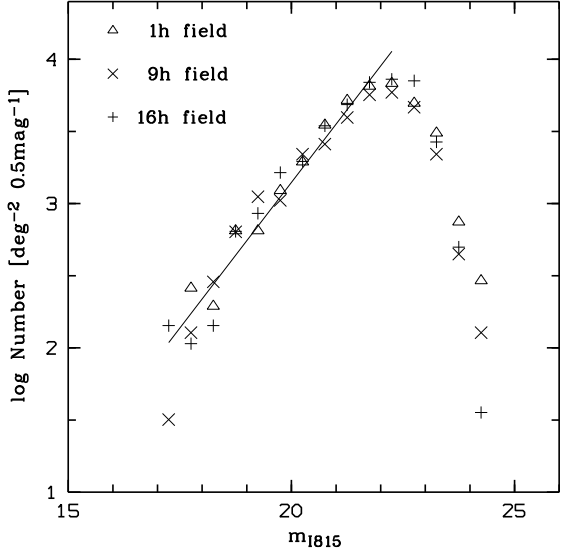}} 
  \caption{Differential number counts in the $I815$ filter. The data points 
    include only those objects which were used in our analysis, i.e. objects 
    which were classified as galaxies and for which a redshift was estimated. 
    The straight line is the fit to these data from $m_{I815} = 17.$ to  $m_{I815} = 22.5$} 
  \label{numcount}
\end{figure}

\begin{figure}
  \resizebox{\hsize}{!}{\includegraphics{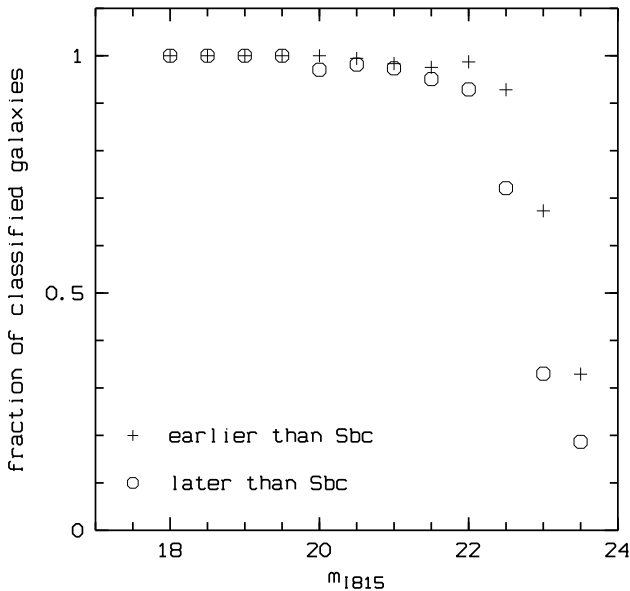}} 
  \caption[]{The completeness of the classification process derived from simulations 
(see sect. \ref{montecarlo}).}
  \label{class}
\end{figure}

In Fig. \ref{numcount} 
we show the differential number counts of objects which are classified 
as galaxies and for which the classification algorithm can determine 
a redshift.  As this figure shows, our data are effectively complete 
to $m_{I815}=22$; only in the faintest 
brightness bin we have to correct for incompleteness. The slope in the number counts 
from $m=18$ to $m=22$ is $0.403 \pm 0.018$ and agrees well with the results of 
 \cite{non}; fitting their data in the same magnitude range we 
obtain a slope of $0.39$. Fig. \ref{class} shows the completeness of the 
classification as function of magnitude derived from the simulations described 
in section 4.

\cite{lilly95} have pointed out the advantages of selecting a sample in the I-band, 
which are (i) good match to B selected local samples since the B band is 
redshifted to I for $ z \approx 0.9$ and (ii) minimization of selection 
effects since k-corrections  are much smaller and depend less on  the SED 
of the object in the I-band than in shorter wavelength filter bands. 
Put another way, surveys that select the objects at shorter wavelengths 
are more affected by selection effects, since at given measured B-magnitude 
galaxies of different morphological types will contribute to the rest frame B band 
luminosity function.

Since the redshifted B-band always falls between two of our filters for 
redshifts $z \le 1.04$, the rest-wavelength B-magnitude is computed for all 
objects by interpolation between the 2 adjacent CADIS filters. 
We emphasize that we do {\em not}  have to derive K-corrections from 
model spectra or polynomial extrapolations, rather we simply interpolate 
measured data.

\begin{figure*}
  \centering
  \includegraphics[width=17cm,clip]{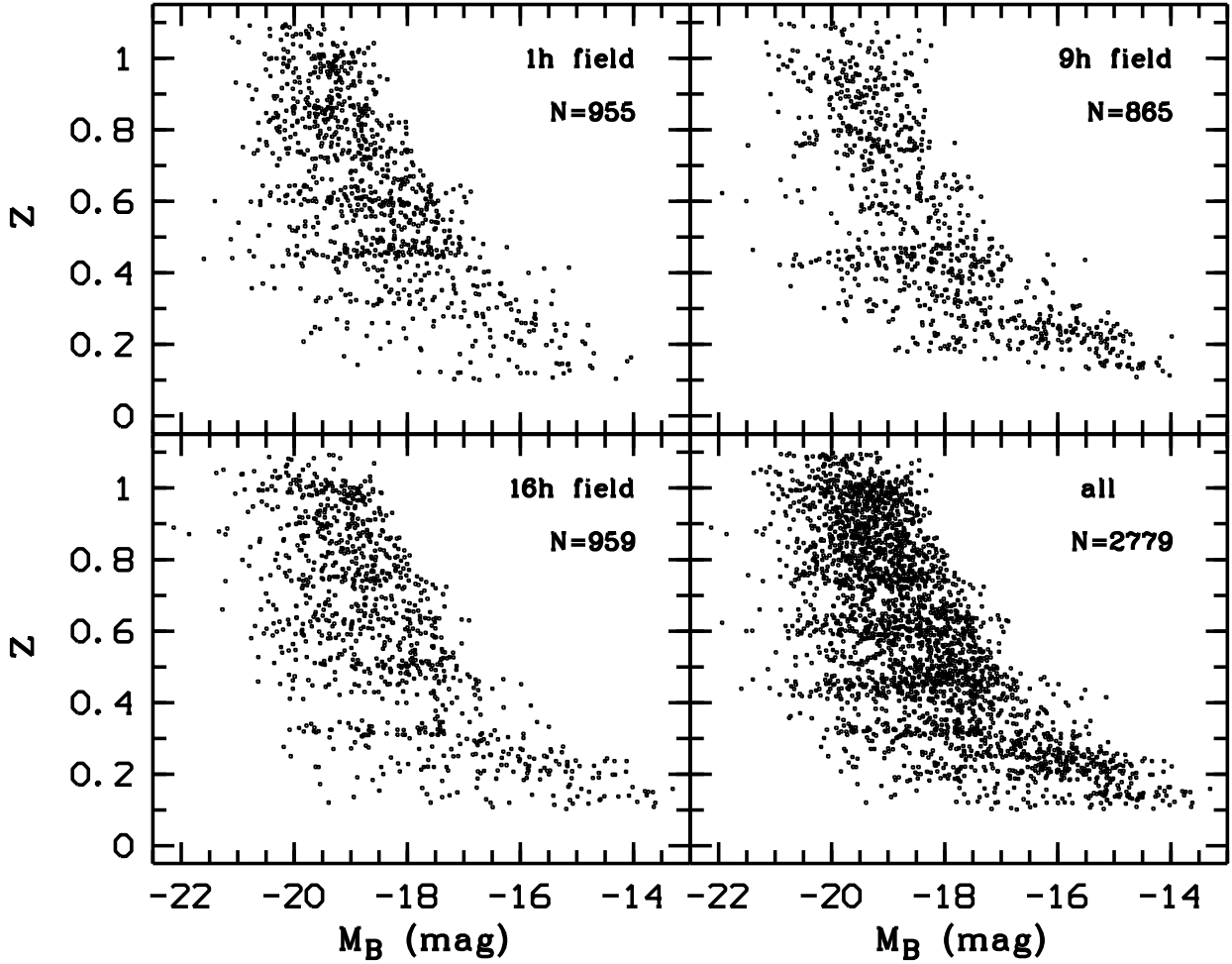} 
  \caption[]{The $M_B-z$ diagrams for the galaxies in our survey for the three fields 
    individually and combined. Absolute magnitudes calculated for a 
    $ H_{0} = 100 \, km s^{-1} Mpc^{-1}, q_{0} = 0.5$ cosmology.}
  \label{mz-dia}
\end{figure*}

Figure \ref{mz-dia} shows our basic observational result, the 
location of the galaxies in the $M_B-z$ diagram. For consistency with earlier work 
the absolute B-magnitudes have been computed using  
$ H_{0} = 100 \, km s^{-1} Mpc^{-1}, q_{0} = 0.5$. 
Note that at $z=1$ the inferred absolute magnitudes would be 0.6 mag brighter for an 
$H_{0}= 100 \, km s^{-1} Mpc^{-1}, \Omega_m = 0.3, \Omega_{\Lambda} = 0.7$ cosmology.

The upper right portion of this M-z diagram is devoid of galaxies which reflects our 
magnitude limit; since this limit is set to $m=23.0$ in the $I_{815}$ filter, 
it appears as a sharp cutoff in the $M_B$ magnitudes only around $z \approx 0.8$. 
The region $ z \le 0.3$ is sparsely populated which is due to 
the relatively small field of view covered by our data; 
we have therefore concentrated our analysis to $z \ge 0.3$.  
The distribution of the galaxies in Fig. \ref{mz-dia} shows conspicuous 
horizontal bands; all three fields combined show a more 
homogeneous distribution. Inspection of the spatial distribution 
of the galaxies in our fields shows that the $9h$ field 
contains a sheet of galaxies near $z=0.2$. Since the local bin contains 
only very few galaxies when the $9h$ field is taken out, we have therefore 
restricted our discussion to the redshift range $z=0.3-1$. A minor 
contribution to the inhomogeneity  in redshift distribution is caused 
by the special choice of filters, which favors some redshifts. 
However, this can be corrected very well (see sect. \ref{compcorr}).

\section{Determination of the luminosity function}

\subsection{Methodology}

The Schechter function is a convenient way to parametrize the luminosity 
function of galaxies. However, Fig. \ref{mz-dia} illustrates that 
the Schechter function cannot be equally well determined at all redshifts 
as the segment of the luminosity function which is covered by data 
changes with redshift. We would 
expect a poor determination of $M^*$ for $z=0.1-0.3$  since there are few 
galaxies brighter than $M^*$  in our relatively small field of view, 
and a poor determination of $\alpha$ for $z \approx 1$ since there are fewer 
galaxies fainter than $M^*$ due to the magnitude limit of the survey. Since the 
parameters of the Schechter function are covariant, the 
parametric description of the luminosity function may become ill defined. 
Thus it is important to use both parametric and non-parametric 
estimations of the luminosity function.

Several estimators to derive  the luminosity function from a set of M-z data 
have been developed in the past. We used  a modified $V_{max}$ estimator 
as non-parametric estimate of the luminosity function and a modified maximum 
likelihood estimator (STY) as parametric estimate (see eg.  \cite{eep1988}). 
These modifications include an explicit completeness correction, which is given 
by a function $C(z_i,SED_i,M_i)$ which is the probability with which a 
galaxy at redshift $z_i$ with $SED_i$ and absolute 
magnitude $M_i$ enters our sample. The determination of C is described 
in Section \ref{compcorr}.

In its standard form, the $V_{max}$ estimator is computed from 
$ \Phi(M) dM = \sum_i \frac{1}{V_i(M)}$ where $V_i$ is the total 
comoving volume in which galaxy {\it i} could lie and still be included 
in the sample. The volume is determined by   the redshift and apparent 
magnitude boundaries, and the sum is taken over all galaxies in the 
$ [ M-dM/2, M+dM/2 ]$ interval and within a redshift interval 
$ [ z_{min}, z_{max} ] $.  
We compute the volume from 

\begin{equation} \label{vmax}
 V_i = {d\omega} \int_{z_{min}}^{z_{max}} C(z,SED_i,M_i) \frac{dV}{d\Omega dz} dz
\end{equation}

where $dV/(d\Omega dz)$ is the comoving volume 
element per unit solid angle measured in steradians and unit redshift, 
and $d\omega$ is the area in steradian covered by the survey. 
Thus the completeness function not only corrects for detection incompleteness, 
but also sets the limits of the integral since $ C(z_i,SED_i,M_i) = 0 $ 
if an object lies outside of the survey criteria. 
The square roots of the variances  $var(\Phi) =  \sum_i \frac{1}{{V_i(M)}^2} $ 
are used as $1 \sigma$ error bars on the $V_{max}$ data points.

In  the STY method the probability $p_i$ of observing galaxy i is 
given by $ p_i dM = \Phi(M_i) dM / \int_{-\infty}^{M_{lim(z)}} \Phi(M) dM $ 
where $\Phi(M)$ is the luminosity function and $M_{lim(z)}$ is the absolute 
magnitude corresponding to the magnitude limit of the survey  for the redshift 
of the galaxy. The completeness correction has been incorporated  
by replacing $\Phi(M_i)$ with $\Phi(M_i) C(z_i,SED_i,M_i)$ so the likelihood 
to observe all galaxies is given by

\begin{equation} \label{sty}
 L = \prod_{i=1}^{N_{obs}} \frac{\phi(M_i) C(z_i,SED_i,M_i)}{ \int_{-\infty}^{+\infty} \Phi(M) C(z_i,SED_i,M) dM}
\end{equation}

where $N_{obs}$ is the total number of galaxies in the (sub)sample.
Here again the completeness function C also sets the limits of the integral.
The best parameters $\alpha$ and $M^*$ are found by maximizing $\ln  L $.  
Errors in these paramaters are derived from the likelihood  contours 
at $\ln L - \frac{1}{2} \Delta\chi^2$ where $ \Delta \chi^2$ is the change in 
$\chi^2$ corresponding to the desired confidence level for 2 degrees of 
freedom. It should be noted that these errors underestimate the true ones, 
because they do not account for the  cosmic inhomogeneities apparent from Fig. \ref{mz-dia}. 
Since the normalization factor $\Phi^*$ cancels out in equation \ref{sty} 
we have determined it from  $\Phi^* = N_{obs}/N_{pred}$ where 

\begin{equation} 
N_{pred} = {d\omega} \int_{z_{min}}^{z_{max}} \int_{-\infty}^{+\infty}  \frac{dV}{d\Omega dz} \Phi(M) C(z,M) dz dM
\end{equation}

is the predicted number of galaxies using the derived parameters 
$\alpha$ and $M^*$ in the luminosity function $\Phi(M)$, 
and $ C(z,M)$ is the mean of C over all SEDs. 
The error of $\Phi^*$  was determined through bootstrapping.

\subsection{Completeness Correction} \label{compcorr}

The completeness function C was determined as follows. 
For a given set of colors and their errors, the color classification algorithm 
(Wolf et al. 2000) estimates redshift and SED for each object. After this 
classifying process an object drawn from a given point 
($SED_i,z_i,M_{B_i}$) in the 3 dimensional parameter 
space will appear at $(SED_i',z_i',M'_{B_i})$. Even for random measurement 
errors the expectation value of 
this latter point in the parameter space is not 
necessarily identical to the original one, so the observed parameter 
space may have systematic over- or underdensities introduced by the classification 
process. Furthermore, the object may not be classified at all 
which indeed sets the completeness limit to our data. 
However, both completeness limit and possible classification systematics
 can be effectively calculated and corrected by 
Monte Carlo simulations of the observations.

From the galaxy template library we have constructed a grid of 625191 input 
``objects'' brighter than $m_{I815} = 23.5$ in the 3 dimensional parameter space  
$SED-z-M_{B}$ where $0 < SED <100$, $ 0 < z < 1.1$ , $-12 < M_B < -24$. 
Apparent magnitudes were calculated for each object corresponding 
to its redshift and Gaussian noise was added, matching  the real 
measurement errors. Then  this catalogue of simulated objects was 
passed  through the classifying process and a new $M_{B_i'}$ using the 
redshift $z_i'$ estimated by the classifying routine was calculated. 

Dividing the $SED-z-M_{B}$ parameter space into  cells, 
the  completeness function $C(z,SED,M_B)$ in each cell 
is obviously given by the ratio objects classified to be in that cell 
${N_{SED_i',z_i',M_{B_i'}}}$ to  objects initially in it: 

\begin{equation} \label{comp}
C(z_i,SED_i,M_i) = \frac{N_{SED_i',z_i',M_{B_i'}}}{N_{SED_i,z_i,M_{B_i}}} .
\end{equation}

The parameter space was devided into  25500 cells. 

\begin{figure} 
  \resizebox{\hsize}{!}{\includegraphics{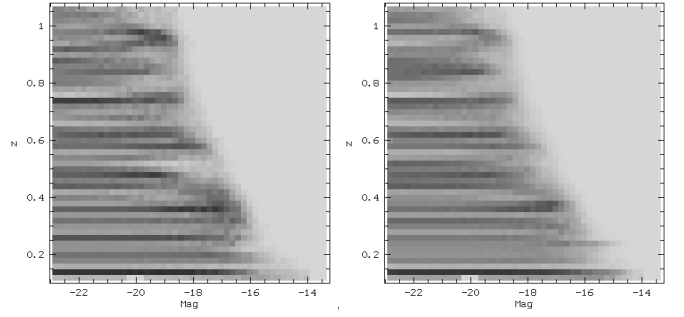}} 

  \caption[]{The completeness correction function C for galaxies earlier than Sbc(left) 
    and galaxies later than Sbc (right). The grey scale coding is linear from 
    C=0 (white) to C=2 (black).}
    \label{compl1}
\end{figure}

The determination of C was done separately for the two cosmologies; 
in Fig. \ref{compl1} we show  2 D averages  for $SED= 0-50 $ and $SED= 50-100$ for the 
$\Omega_m=1, \Omega_{\Lambda}=0$ cosmology.  
It is obvious that the classification process favors 
some regions of the redshift space and discriminates against others; 
this is probably  caused by the special choice of the wavelengths of the  
filters we are using. However, since our definition of C allows 
$ C > 1$, this effect is fully taken into account in the corrections for 
$V_{max}$ (Equation \ref{vmax}) and in the maximum likelihood estimation 
(Equation \ref{sty}).

\section{Monte Carlo Simulations} \label{montecarlo}
The effect of photometric errors on the luminosity functions was studied by  
Monte Carlo simulations. An advantage of our purely photometric classification 
is the possibility to simulate the complete process, including the  redshift assignments  
to the objects.  Though the errors 
of our photometric redshifts are on the order of $\delta z =0.04$ and thus 
larger than typical spectroscopic redshift errors, one would expect 
that their effect is small since they lead to errors in absolute 
magnitude $\delta M = 0.08$ which are much less than the 1-mag bins 
we are using for calculating the luminosity function. Nevertheless, 
we have examined the effect of redshift errors  by numerical simulations. 
%Additionally, these simulations allow to determine systematic effects.  

Simulated galaxies were distributed in the M-z plane with a redshift 
 distribution that corresponds to a constant comoving volume density and 
with absolute magnitudes drawn from a Schechter function with 
$M^* = -19.5, \alpha = -1.0$ for all types of galaxies. 
The SEDs were randomly distributed in the interval $[0, 100]$. Apparent 
magnitudes were calculated using the $\Omega_m=1, \Omega_{\Lambda}=0$ cosmology 
and noise added to the color indices according to 
the actual measurement process for all artificial galaxies. These objects were 
then run through the classifying process. For $m_{I815} > 22.0$ $94 \% $ 
of the input galaxies were classified as such, 
and $86 \%$ of the galaxies have redshift errors $ \le \pm 0.05$. 
In the center of our last magnitude bin $m_{I815} = 22.5$, the corresponding 
numbers are $82 \% $ and $72 \%$, respectively. The classification completeness is 
shown in Fig. \ref{class}. 
In this manner we have formed 5 samples of  galaxies with the number of 
galaxies in each sample roughly as in our data. For these samples we have 
then determined the luminosity function as for the real galaxies.

The agreement between the input and the recovered Schechter parameters is excellent 
in the redshift range $z=0.3-0.9$ where we find displacements in characteristic 
magnitude $ | \delta M^* | \le 0.05$ and slope $| \delta \alpha | \le 0.02$. 
As expected, the parameters are less well recovered for $z=0.1-0.3$ and $z \ge 0.9$:
we obtained $M^* = -19.2$, $\alpha = -0.82$ for $z=0.1-0.3$ 
and  $M^* = -19.84$, $\alpha = -1.31$ for $z > 0.9$. 
The numbers given here apply to   both early and 
late type galaxies; we found no significant dependence on SED type. 

These simulations thus show that we can recover the luminosity function 
parameters very well for $z=0.3-0.9$. The displacements of the recovered 
parameters are smaller than their errors in the actual data. 
For redshifts outside the $z=0.3-0.9$ range the description of the 
luminosity function by the parameters of the Schechter 
function becomes poor.  However, even if the parameters are well 
determined, it may be dangerous to draw conclusions on the 
luminosity function from the parameters alone because they are coupled: 
for example, a decrease in $M^*$ is compensated 
by an increase in $\alpha$ - the error ellipses in Figs. 
\ref{evola} to  \ref{evolc_ac} demonstrate this very clearly 
- and consequently by a decrease in $\phi^*$. Therefore the non-parametric 
form of the luminosity function is indispensable.

\section{Results}

\subsection{The local luminosity function}

\begin{figure}  
 \resizebox{\hsize}{!}{\includegraphics{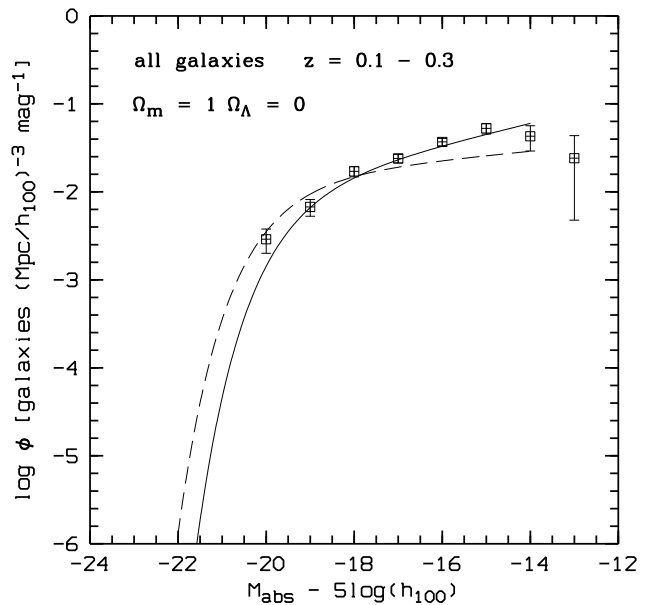}}
  \caption[]{The luminosity function of the 439 galaxies in our sample with 
$z = 0.1-0.3$. The data points were derived from the $V_{max}$ method, the solid line is 
the Schechter function derived from the STY method, the dashed line is the mean of recent 
determinations of the local luminosity function.} 
    \label{local}
\end{figure}

Because of the sheet of galaxies contained in the $9h$ field and the resulting 
problems  (see sect. \ref{sample}), our results for the local redshift bin are 
preliminary until CADIS is completed. Nevertheless we show the luminosity 
function for all galaxies with $z=0.1-0.3$ in Fig. \ref{local}. 
The derived Schechter function is described by 
$\phi^* = 0.0156 \pm 0.004, M^* = -19.2 \pm 0.27, \alpha = -1.30 \pm 0.069$. 
This is agrees  within the errors with an eyefitted  mean of recent  
determinations by \cite{zucca}, \cite{love97}, \cite{mar97}, \cite{mar94a} and \cite{fol} 
which is described by $\phi^* = 0.017, M^* = -19.6, \alpha = -1.15 $.

\subsection{Evolution of the luminosity function}

It has been well established that the local luminosity function is different for 
early and late type galaxies in the sense that the slope $\alpha \approx -0.8$ 
for early type galaxies which are dominated by old populations and 
$\alpha \approx -1.8$ for late type galaxies which are dominated by 
starforming populations (\cite{mar94b}), i.e. the starforming 
galaxy population is  richer in dwarfs. Furthermore, other redshift surveys have 
found a dependence of the evolution of the luminosity function on galaxy type. 
We have therefore divided our sample into three SED bins, E-Sa galaxies with 
$SED \le 30$, Sa-Sc galaxies with $30 < SED < 75$ and starforming galaxies 
with $SED > 75$ (\cite{wolf}, \cite{kinn}). 
We have further divided our sample into three redshift bins  
$z=0.3-0.5$, $z=0.5-0.75$ and $z=0.75-1.04$. 

\begin{figure*}[h]  
  \resizebox{17cm}{!}{\includegraphics{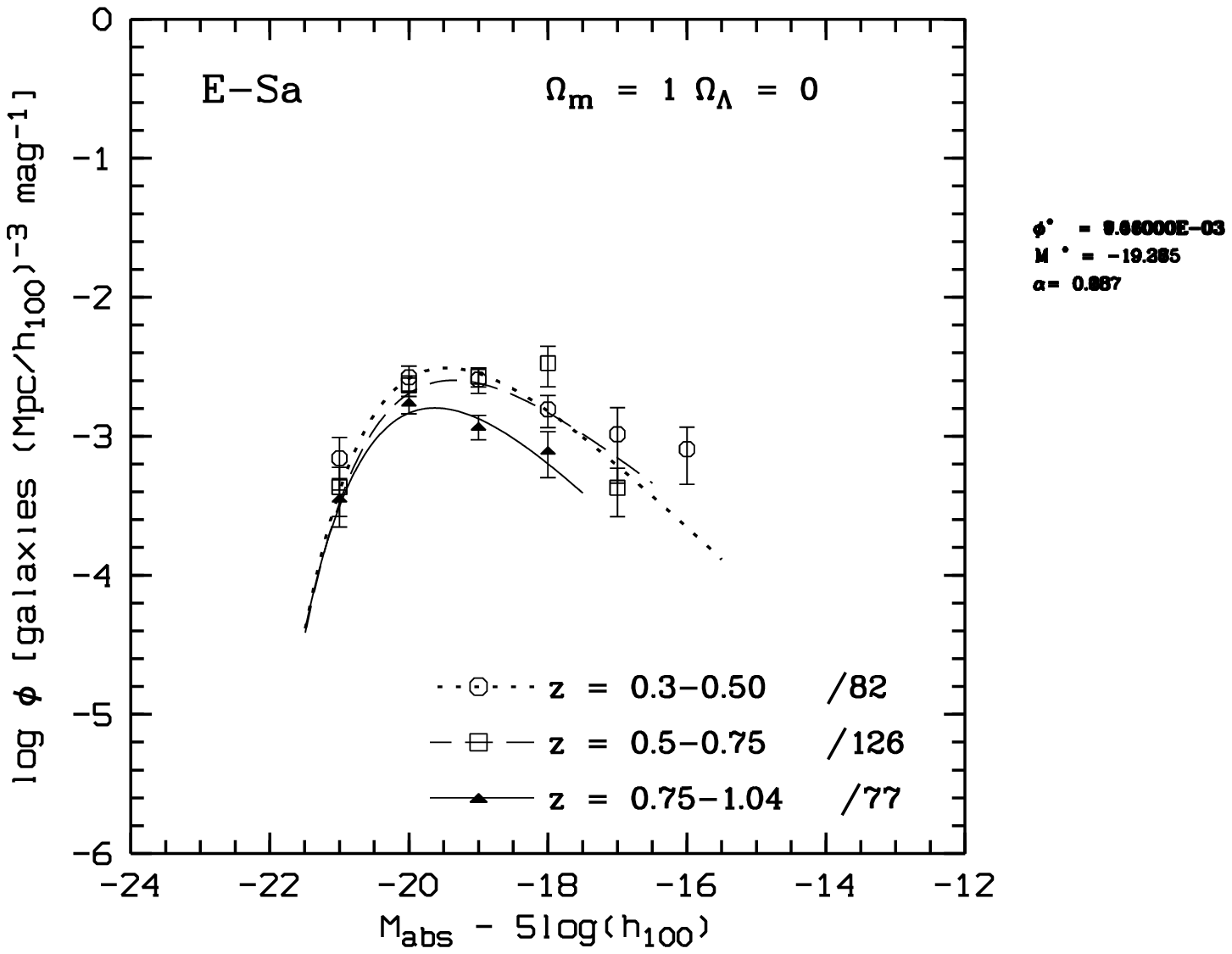},\hspace{1.7cm},
                      \includegraphics{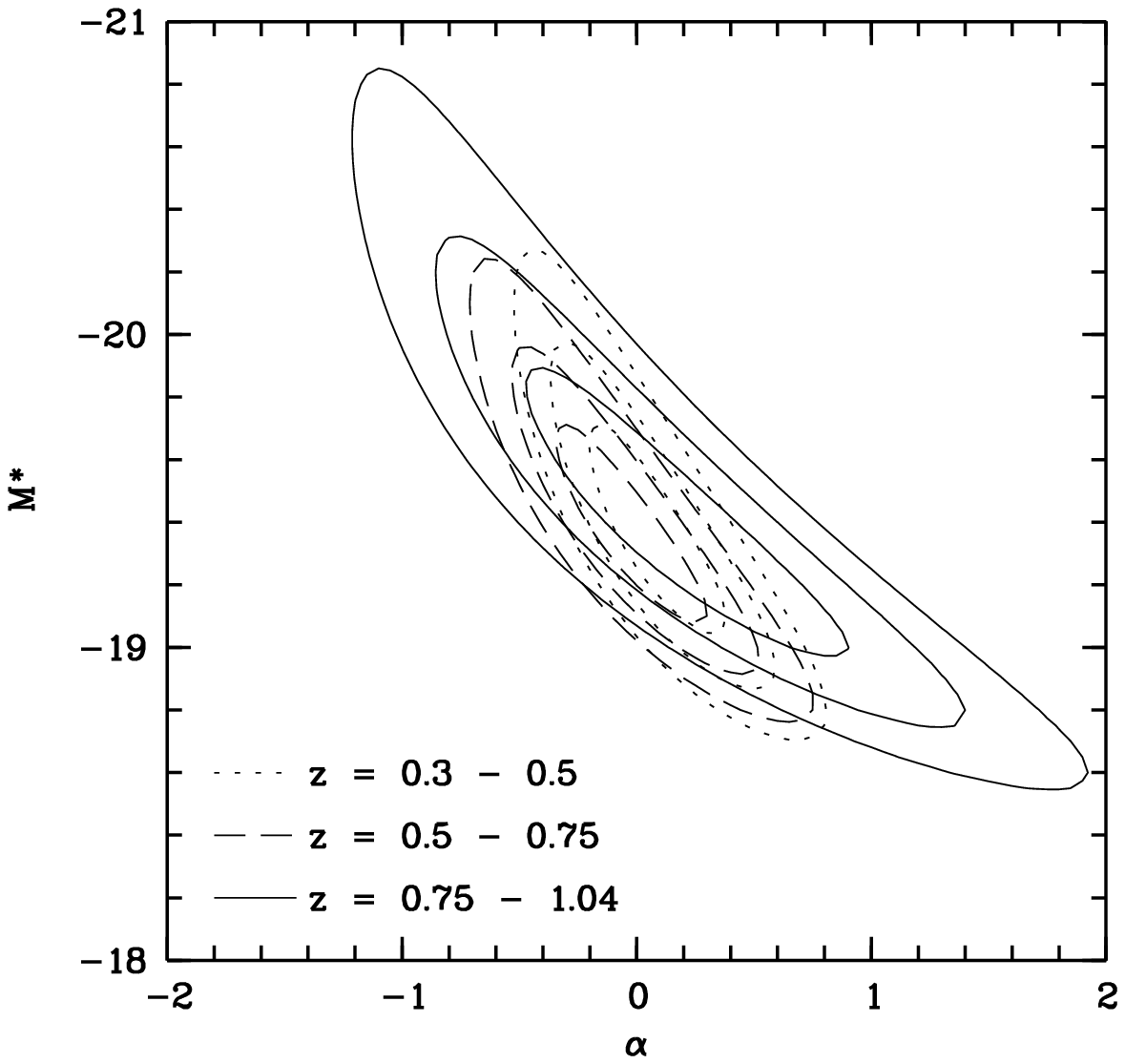}}
  \caption[]{Left panel: The luminosity function of early type galaxies E-Sa  
    for a $H_0=100 \, km s^{-1} Mpc^{-1}, q_0=0.5$ cosmology. $V_{max}$ 
    data points with $1 \sigma$ error bars and Schechter functions derived 
    from the STY method are shown. The numbers are the 
    numbers of galaxies in the corresponding redshift bin. 
    Right panel: $1 \sigma$ and $2 \sigma$ error contours.} 
    \label{evola}
\end{figure*}

\begin{figure*}[h]  
  \resizebox{17cm}{!}{\includegraphics{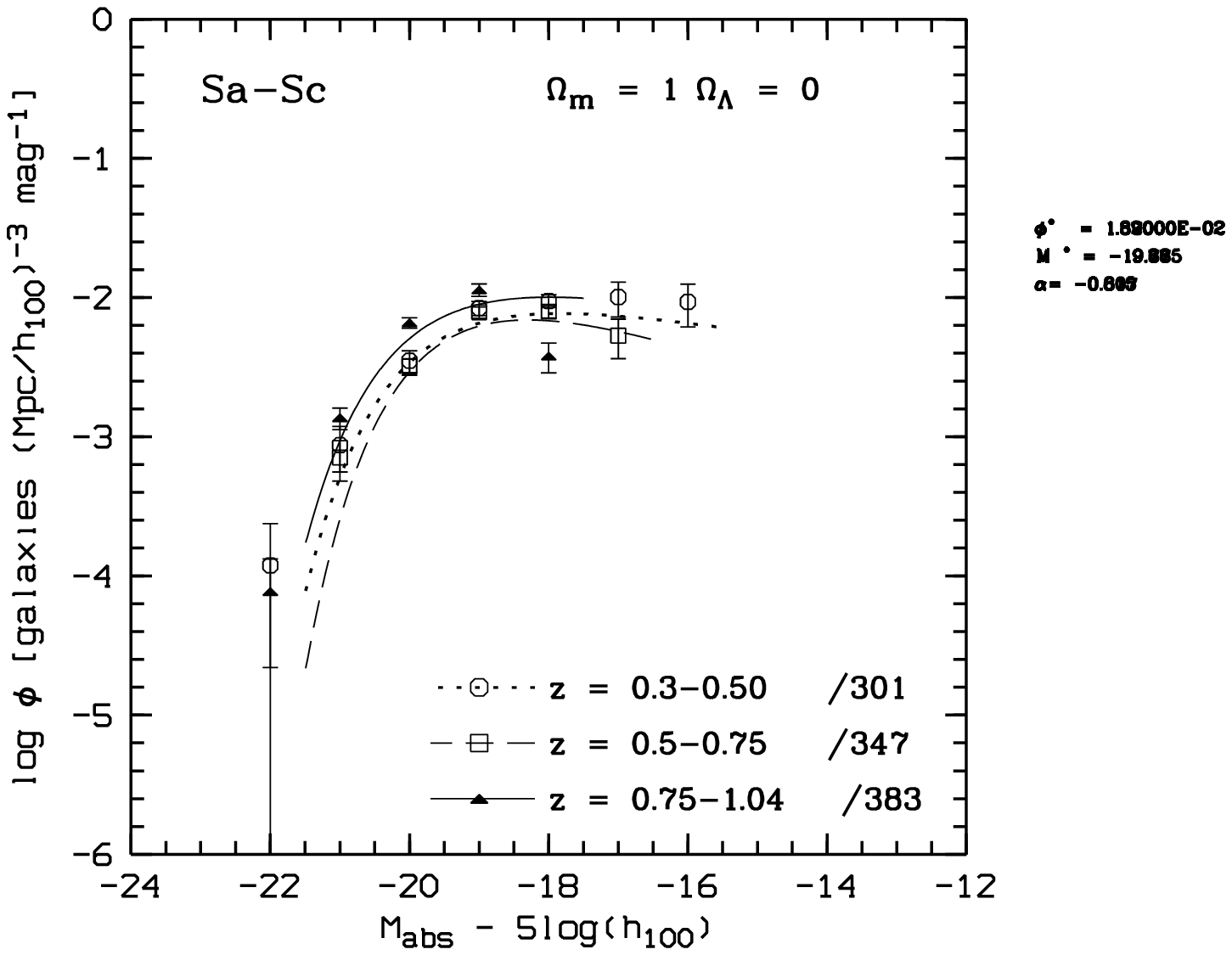},\hspace{1.7cm},
                      \includegraphics{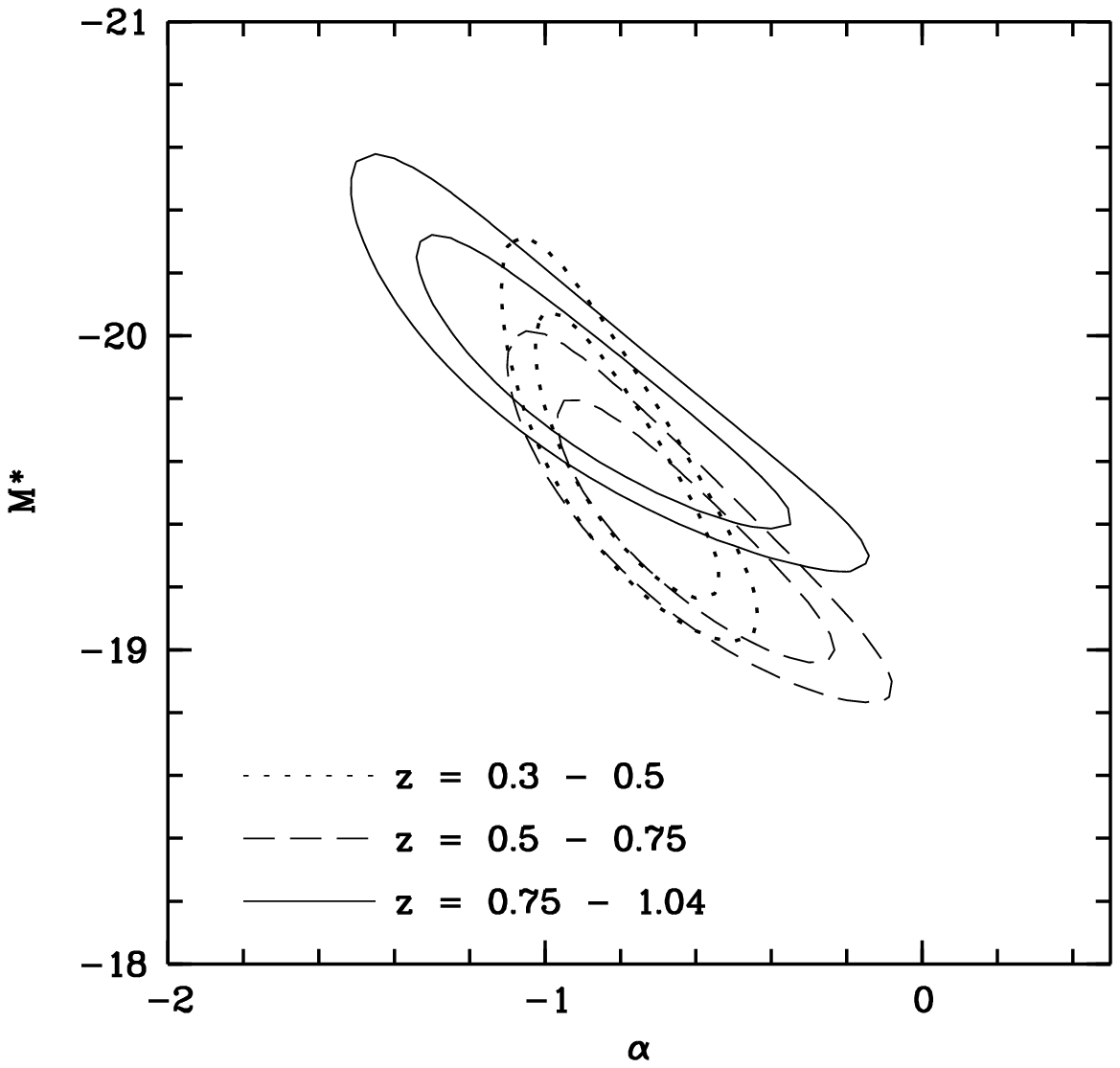}}

  \caption[]{Same as Fig. \ref{evola} but for galaxies of types Sa-Sc.} 

    \label{evolb}
\end{figure*}

\begin{figure*}[h]  
  \resizebox{17cm}{!}{\includegraphics{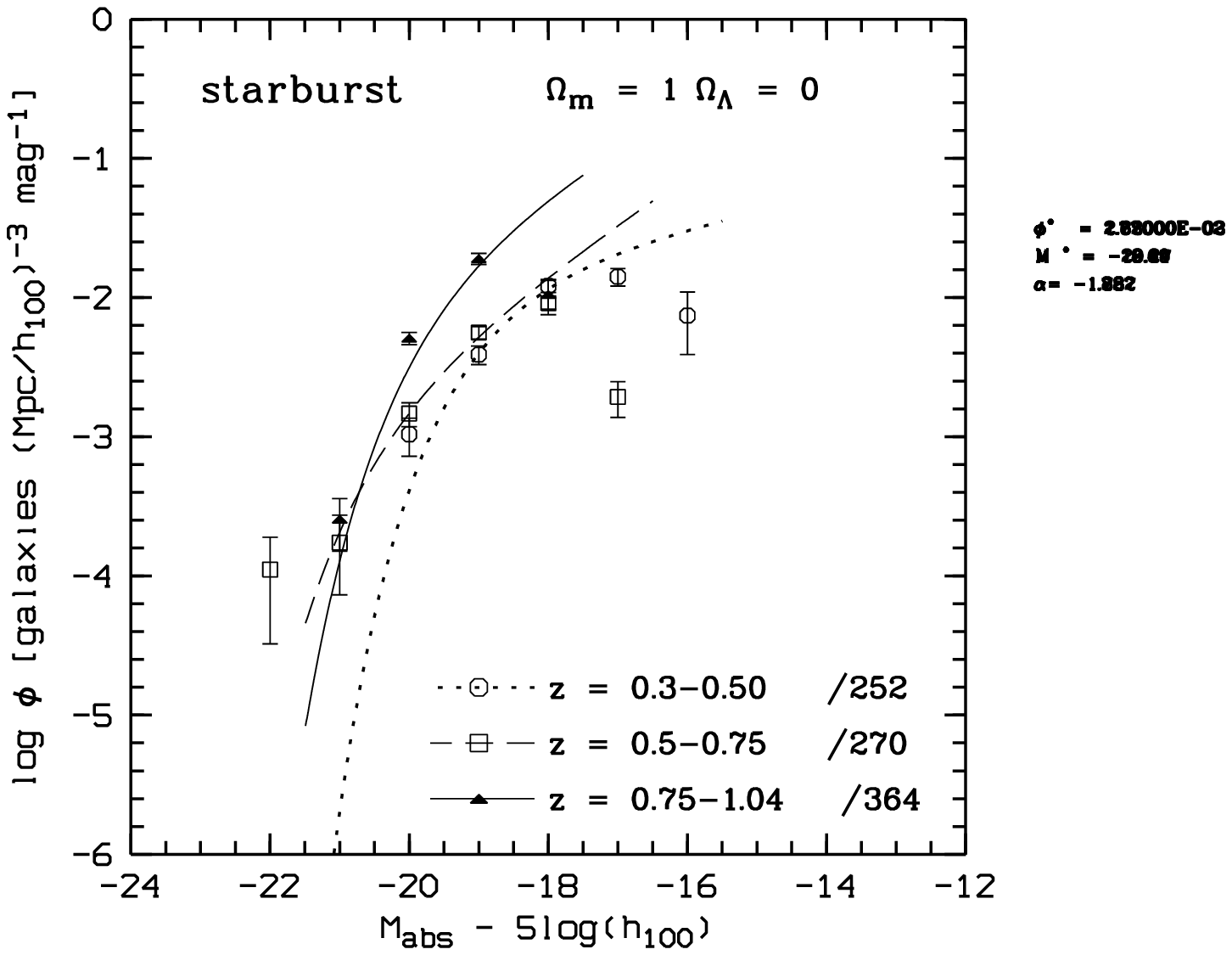},\hspace{1.7cm},
                      \includegraphics{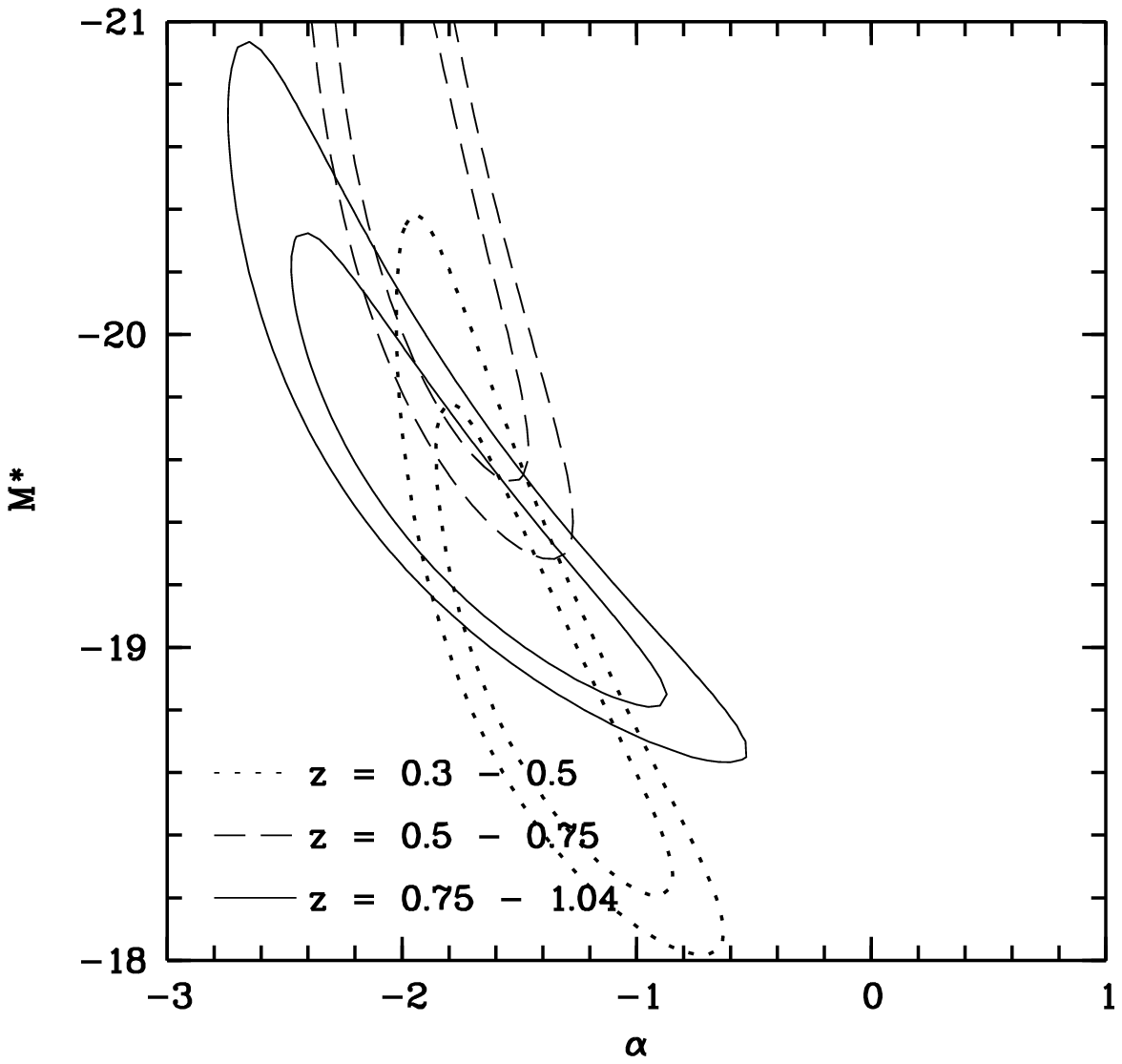}}

  \caption[]{Same as Fig. \ref{evola} but for starbursting galaxies}
    \label{evolc}
\end{figure*}

\begin{figure*}[h]  
  \resizebox{17cm}{!}{\includegraphics{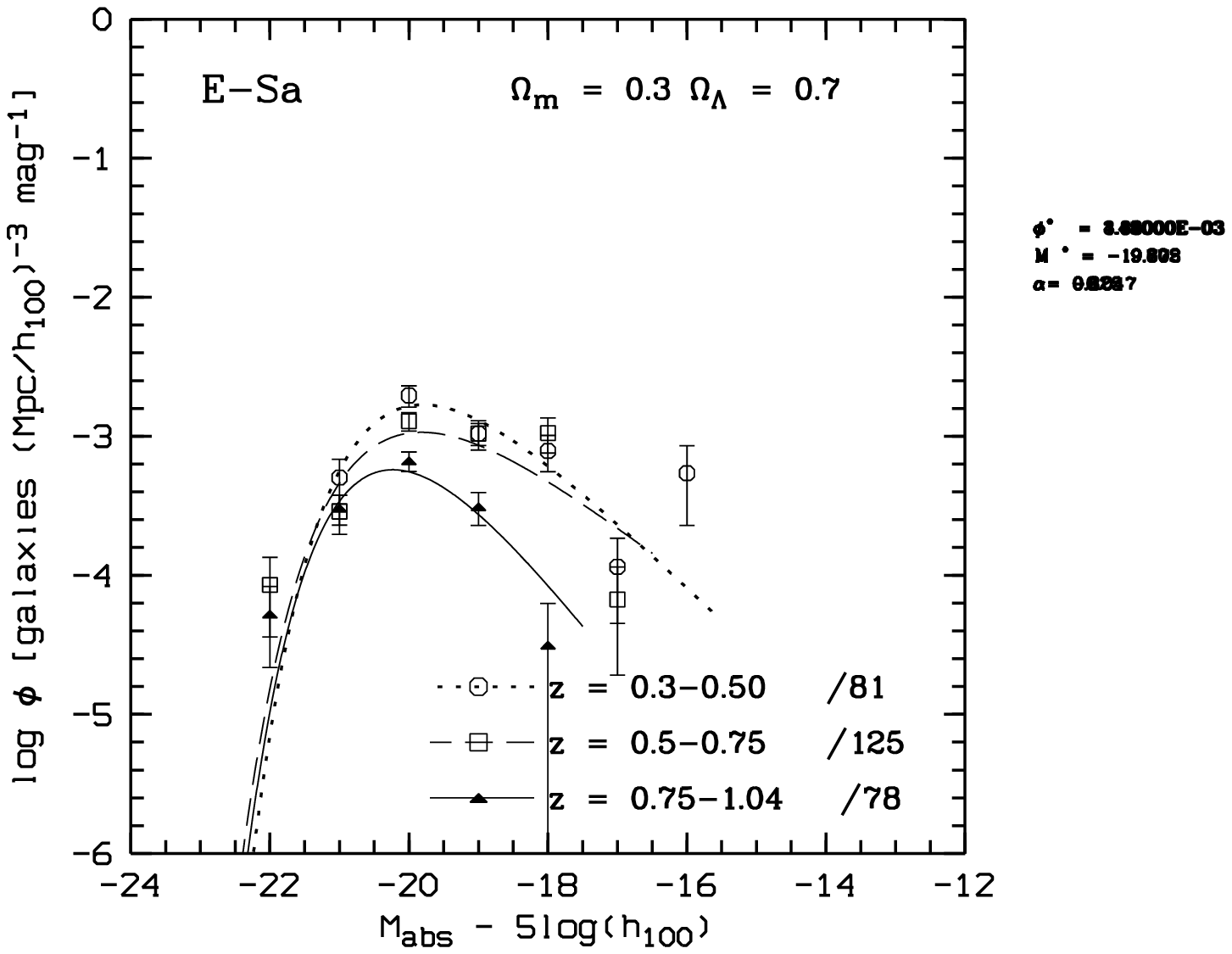},\hspace{1.7cm},
                      \includegraphics{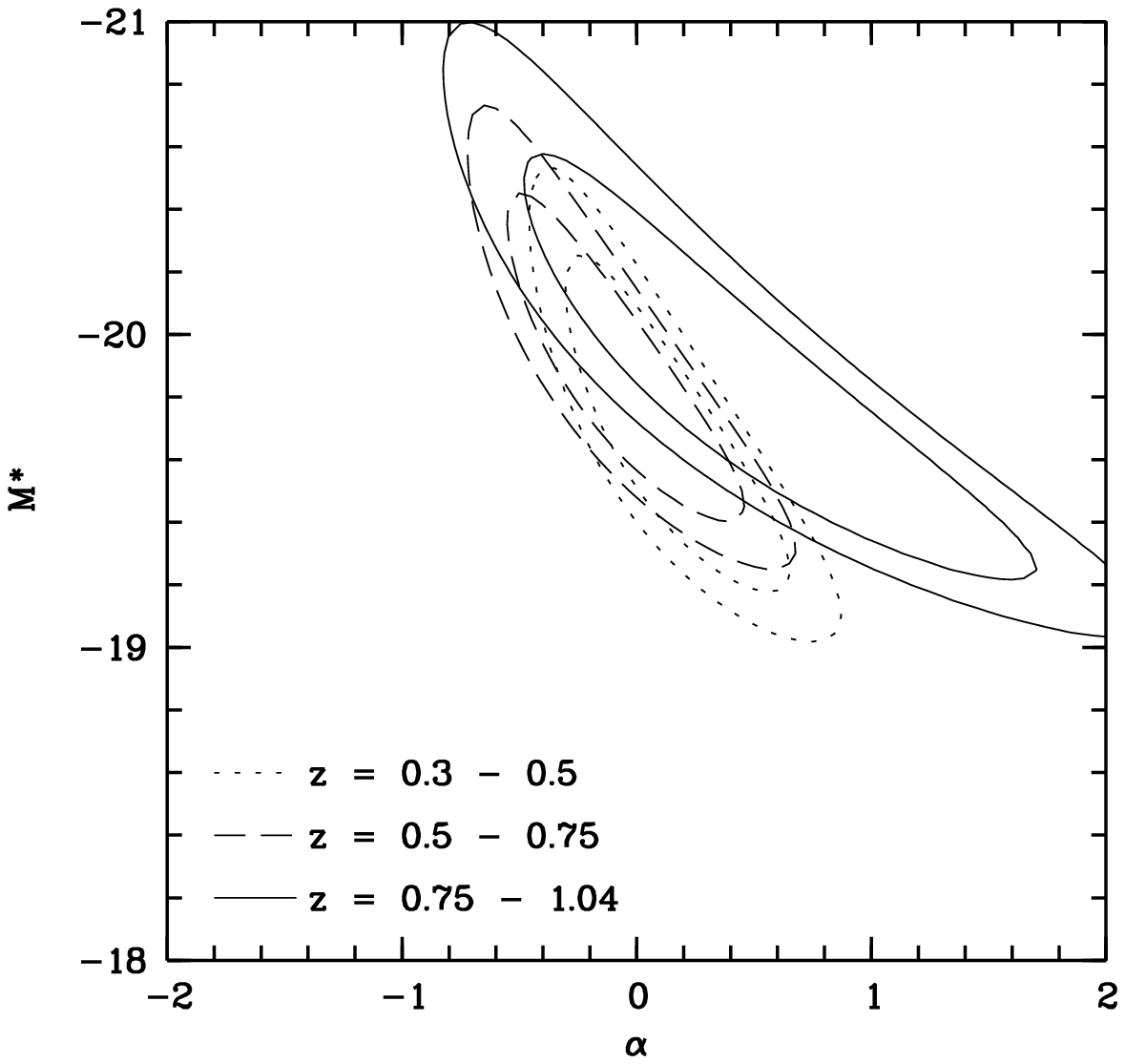}}
  \hfill

  \caption[]{Same as Fig. \ref{evola} but for a $H_0=100 \, km s^{-1} Mpc^{-1}, \Omega_m =0.3, 
    \Omega_{\Lambda}  = 0.7$ cosmology.}

    \label{evola_ac}
\end{figure*}

\begin{figure*}[h]  
  \resizebox{17cm}{!}{\includegraphics{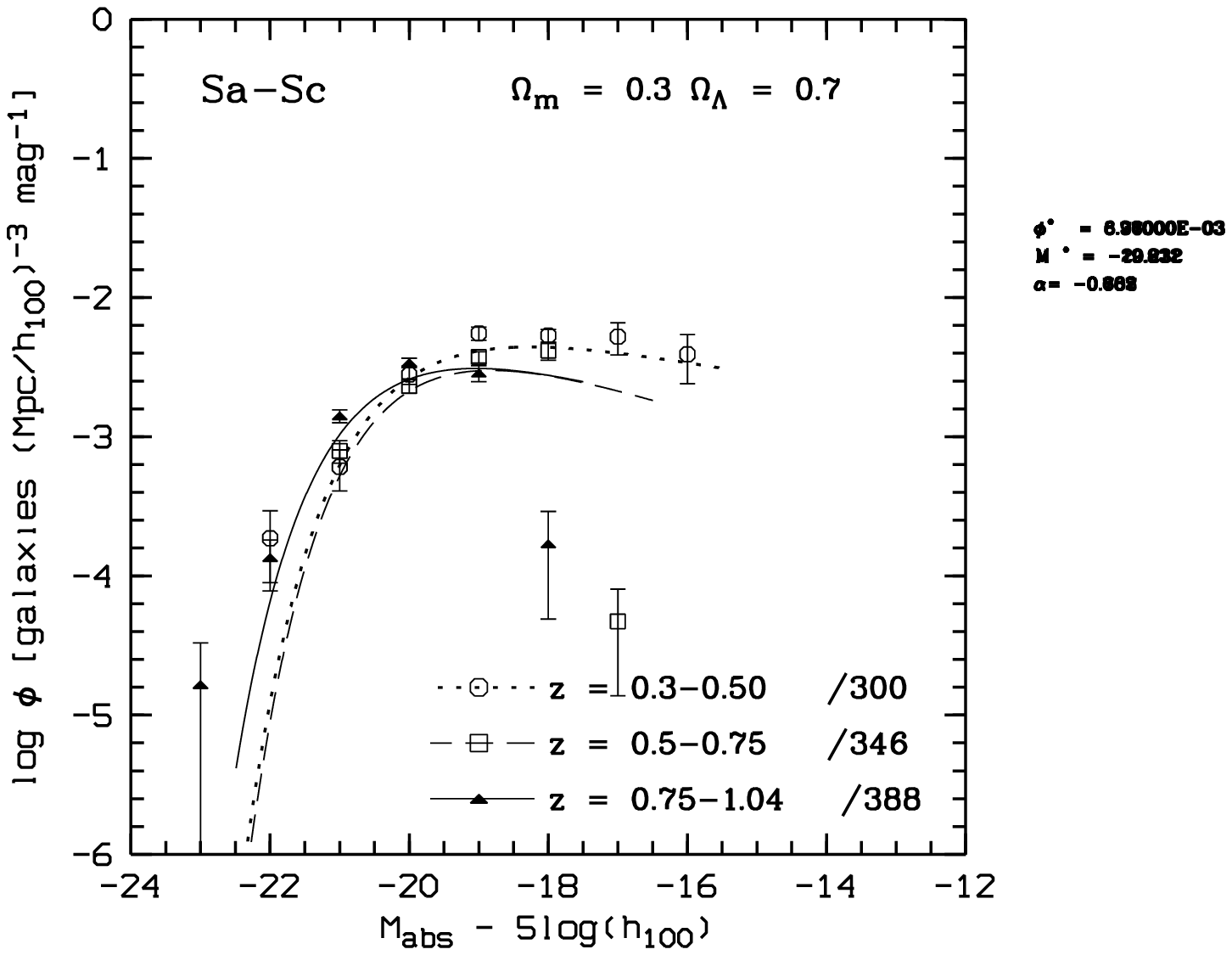},\hspace{1.7cm},
                      \includegraphics{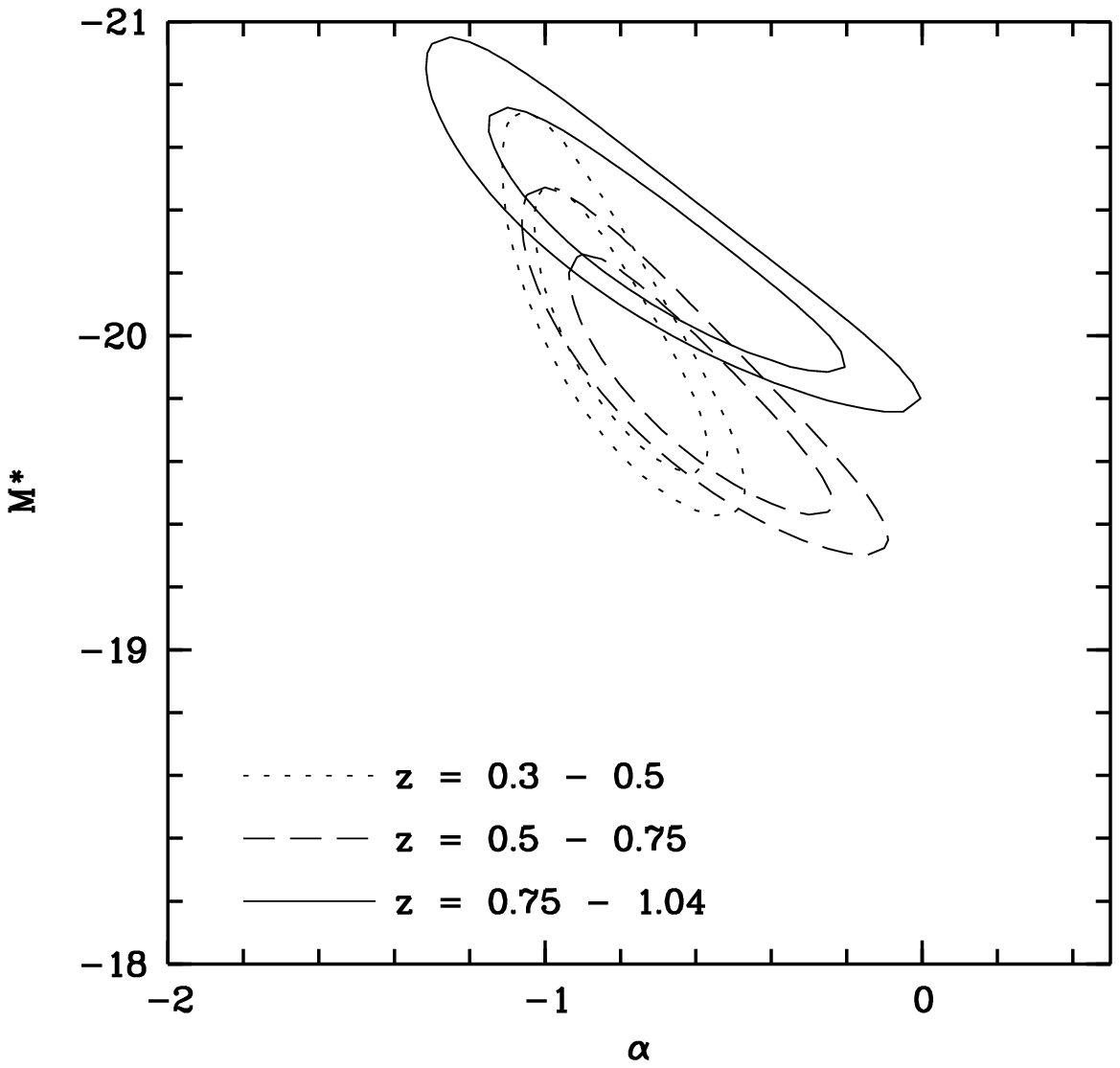}}
  \hfill

  \caption[]{Same as Fig. \ref{evola_ac} but for galaxies of types Sa-Sc} 

    \label{evolb_ac}
\end{figure*}

\begin{figure*}[h] 
  \resizebox{17cm}{!}{\includegraphics{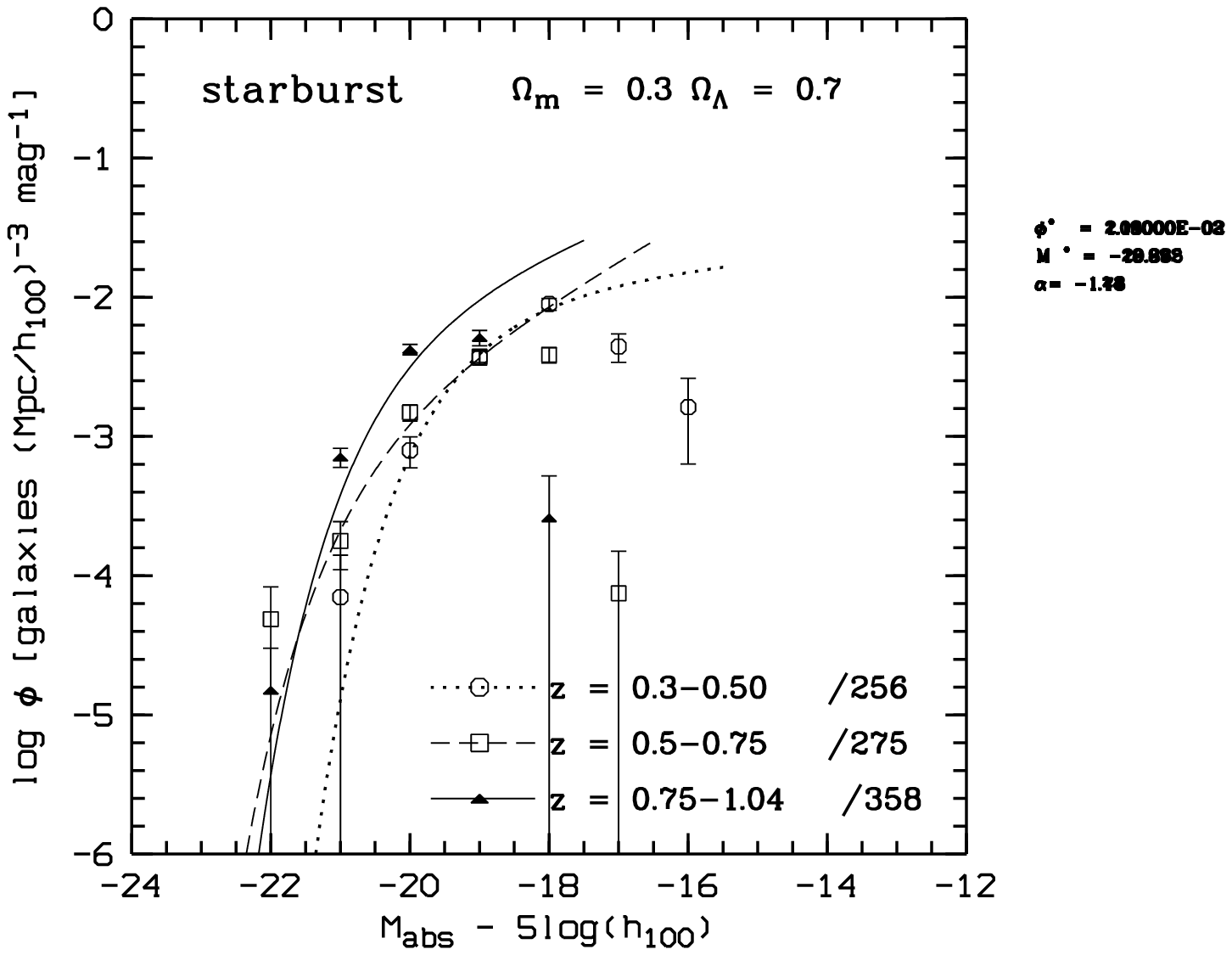},\hspace{1.7cm},
                      \includegraphics{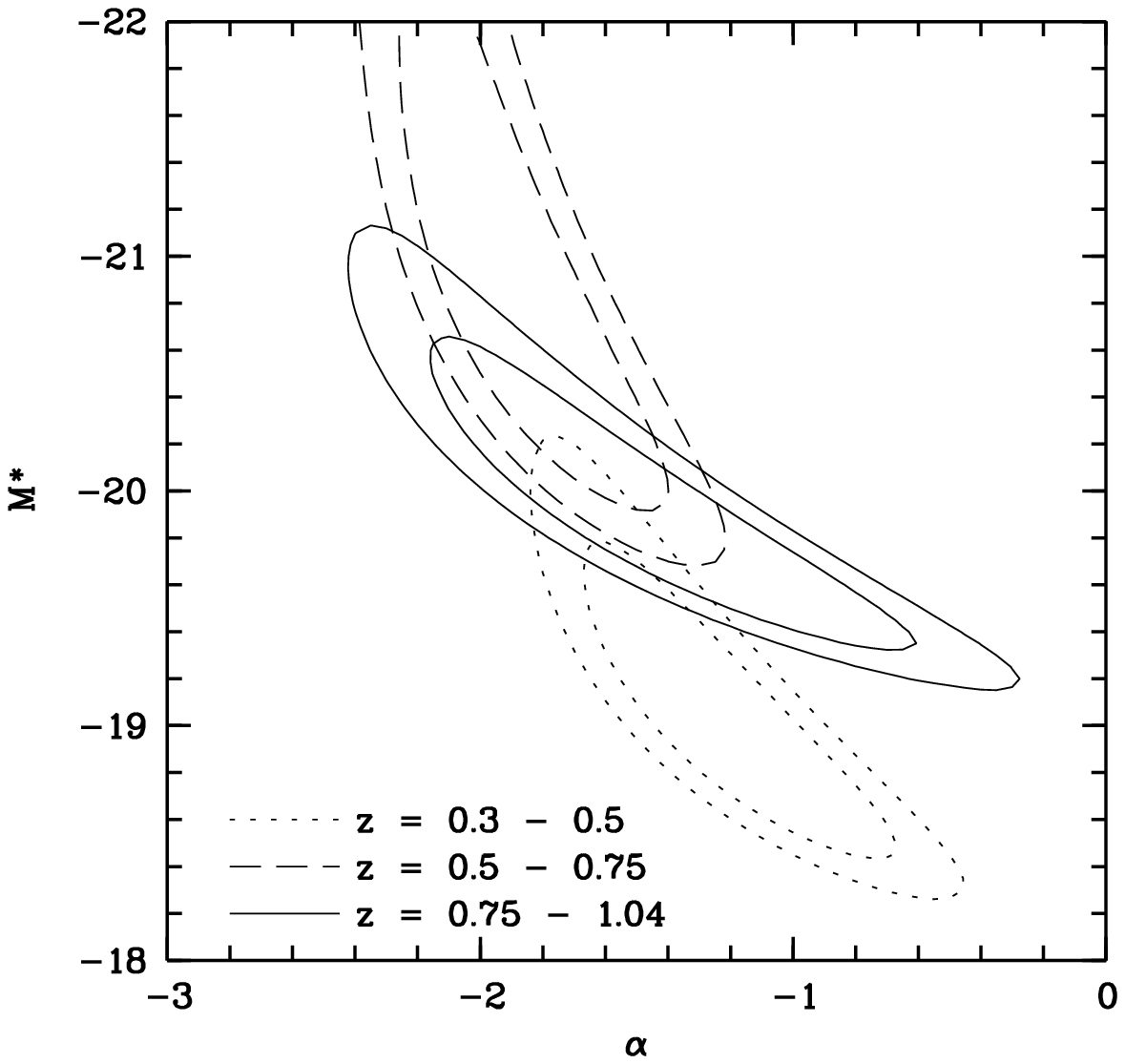}}

  \caption[]{Same as Fig. \ref{evola_ac} but for starbursting galaxies} 
    \label{evolc_ac}
\end{figure*}

To facilitate comparison with other work we have derived the luminosity 
function for an $ H_{0} = 100 \, km s^{-1} Mpc^{-1}, q_0 =0.5 $ cosmology.  
We also give the results for a cosmology with $\Omega_m = 0.3, \Omega_{\Lambda} = 0.7$; 
in order to show the effects of cosmic acceleration in a transparent 
manner, we have again used $ H_{0} = 100 \, km s^{-1} Mpc^{-1}$.  
For $ H_{0} = 62 \, km s^{-1} Mpc^{-1}$ the conversion relations are 
$M_{62} = M_{100}-1.02$ and $\phi^*_{62} = \phi^*_{100} \times 4.2$. 
Since the effects of the cosmology on the luminosity functions and comoving 
space densities are mild, we concentrate the discussions on the $q_0 = 0.5$ cosmology. 
The results are summarized in Figs. \ref{evola} to  \ref{evolc_ac} 
and table \ref{tab_lumfun}.

As Figs. \ref{evola} - \ref{evola_ac} show, the luminosity 
functions of early type galaxies E-Sa have indistinguishable $M^*$ and 
$\alpha$ whereas $\phi^*$ decreases with redshift. For the Sa-Sc 
galaxies  $M^*$ and $\alpha$ are constant, too, but $\phi^*$ is 
increasing with redshift. Note that  the luminosity function 
of the starbursting galaxies with $z=0.75-1.04$ depends sensitively on a 
few bright galaxies in the $M=-22$ bin. Nevertheless, a steepening of 
the luminosity function with redshift is clearly evident.

\begin{figure*}[h]
  \resizebox{17cm}{!}{\includegraphics{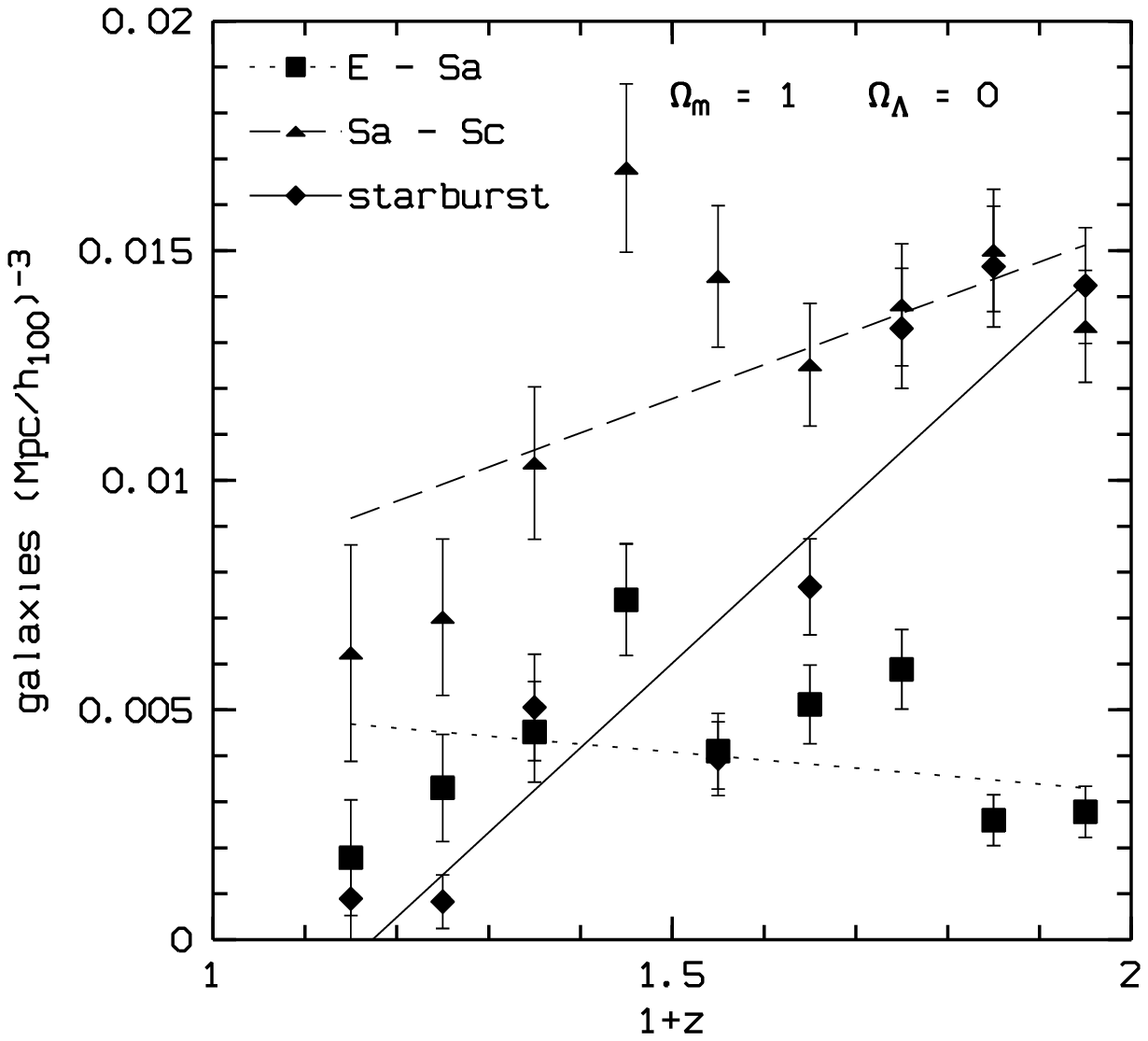},\hspace{1.0cm},
                      \includegraphics{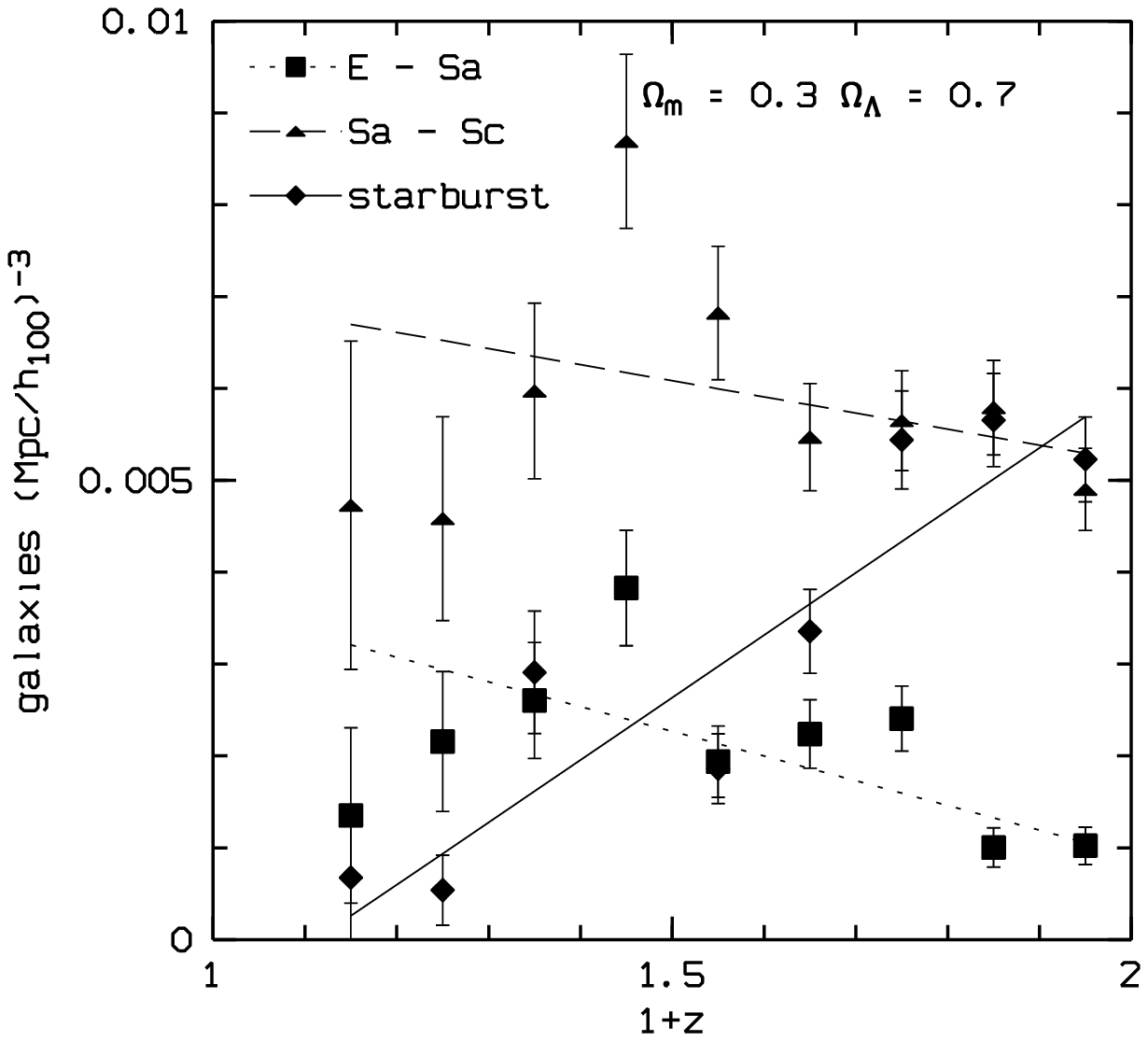}}
    \caption[]{The evolution of the comoving number density  of galaxies $M_B < -18.5$  
      for the two cosmologies. Note the different scaling of the ordinates.}
      \label{densevol}
\end{figure*}

\begin{figure*}[h]
  \resizebox{17cm}{!}{\includegraphics{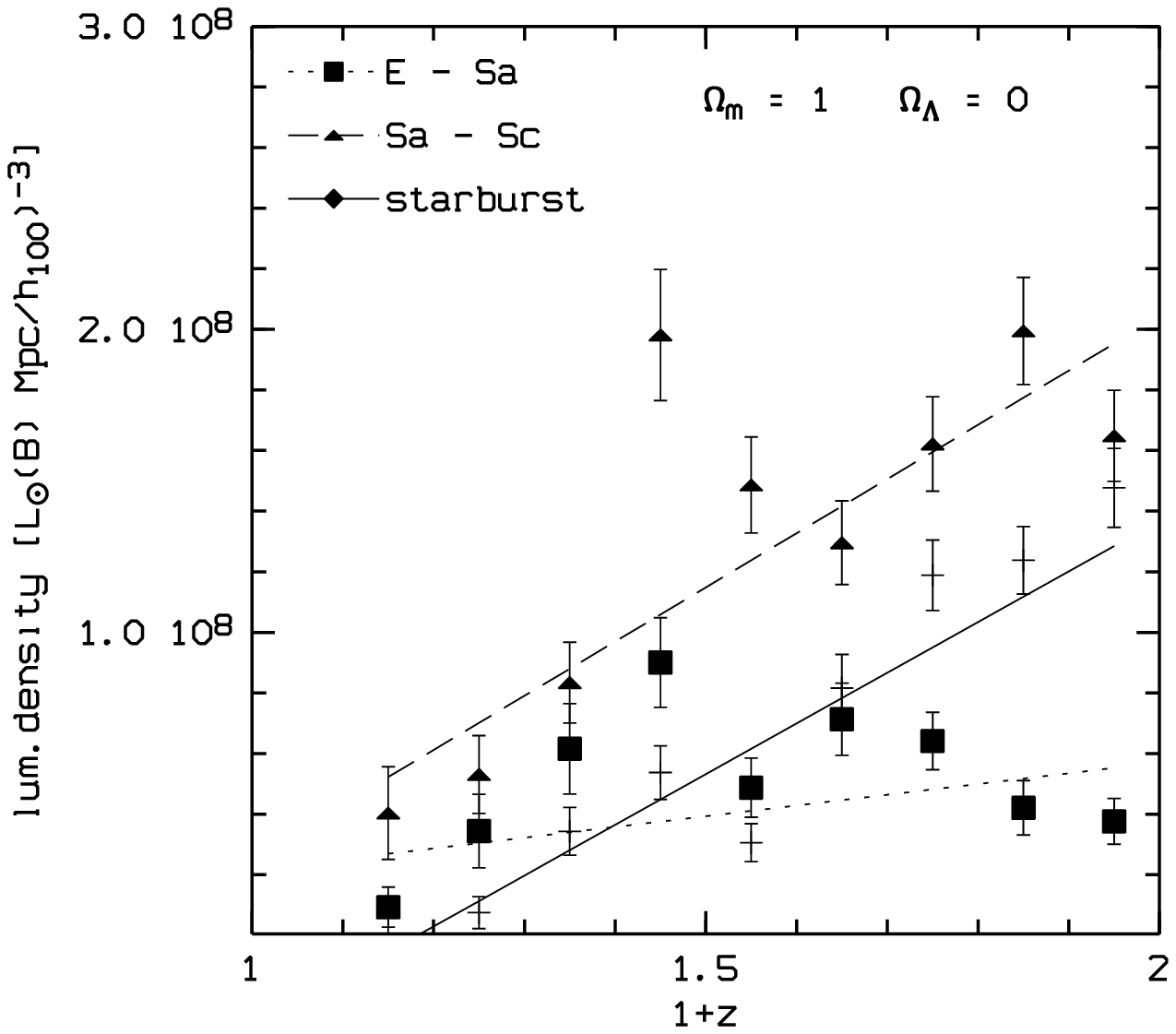},\hspace{1.0cm},
                      \includegraphics{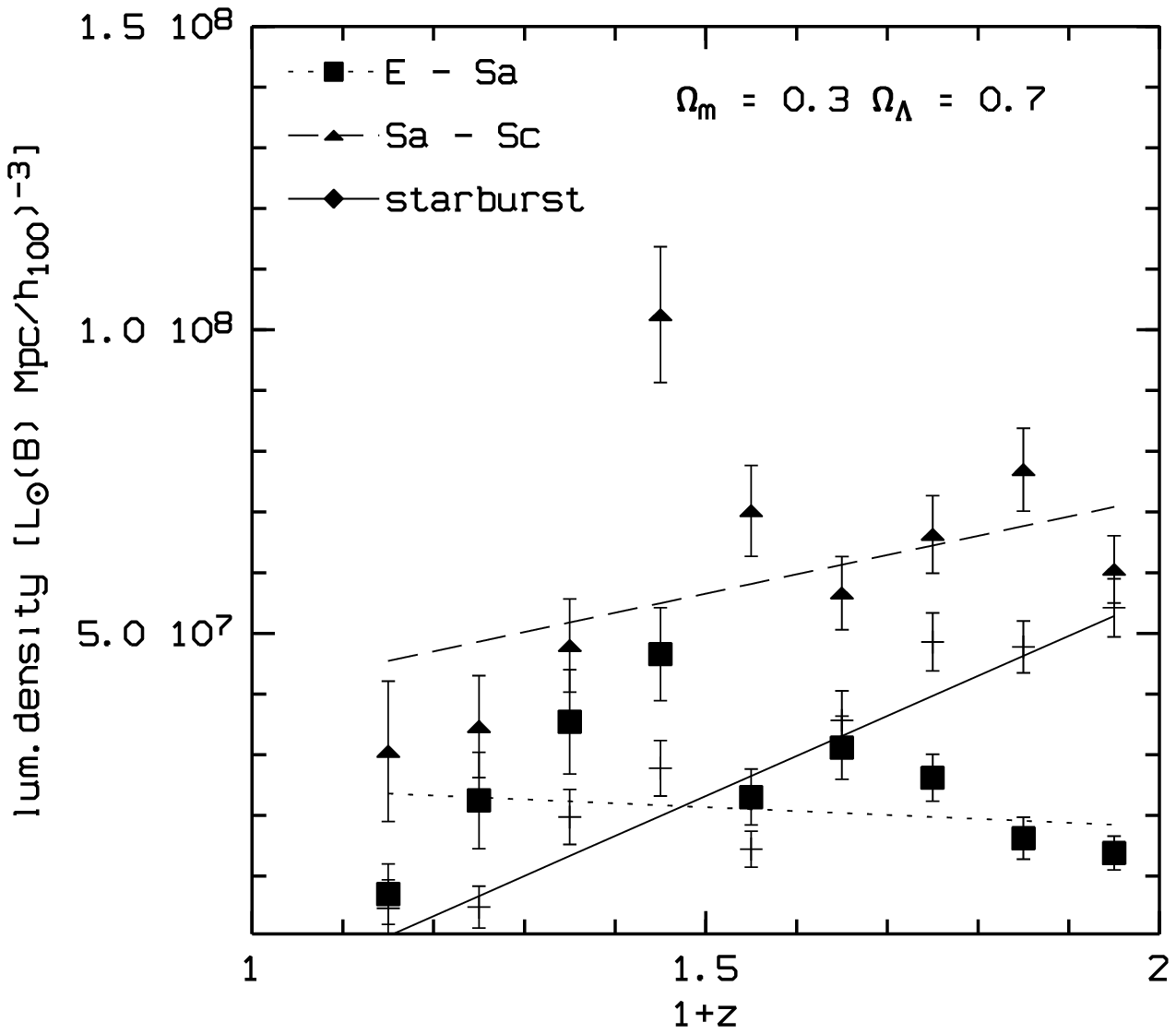}}

    \caption[]{The evolution of the comoving B-band luminosity density  of 
      galaxies $M_B < -18.5$  for the two cosmologies. Note the different scaling 
      of the ordinates.}
      \label{lumevol}
\end{figure*}

To avoid the covariances in the parametric form of the luminosity function,  
we have also analysed the comoving galaxy space density  
and luminosity density as function of redshift in a way which is completely 
independent of the luminosity function. Since our data are 
virtually complete to $M_B = -18.5$ (see Fig.\ref{mz-dia}) 
for all redshifts, we have integrated the numbers and 
luminosities, respectively, for galaxies brighter than this limit 
without completeness correction.  The resulting comoving densities 
are shown as function of redshift in the Figs. \ref{densevol} and \ref{lumevol} 
and the parameters of linear fits to these data are given in table \ref{evol}.

For the $q_0 = 0.5$ cosmology, we find a decreasing comoving space 
density of the early type galaxies. The space density increases with 
redshift for the  Sa-Sc galaxies as well for the starburst galaxies. 
The behaviour of the B-band comoving luminosity density is similar. 
For an $\Omega_m =0.3, \Omega_{\Lambda}  = 0.3$ 
cosmology, the trends in the data are similar but with reduced gradients.

\begin{figure*}[h]
  \resizebox{17cm}{!}{\includegraphics{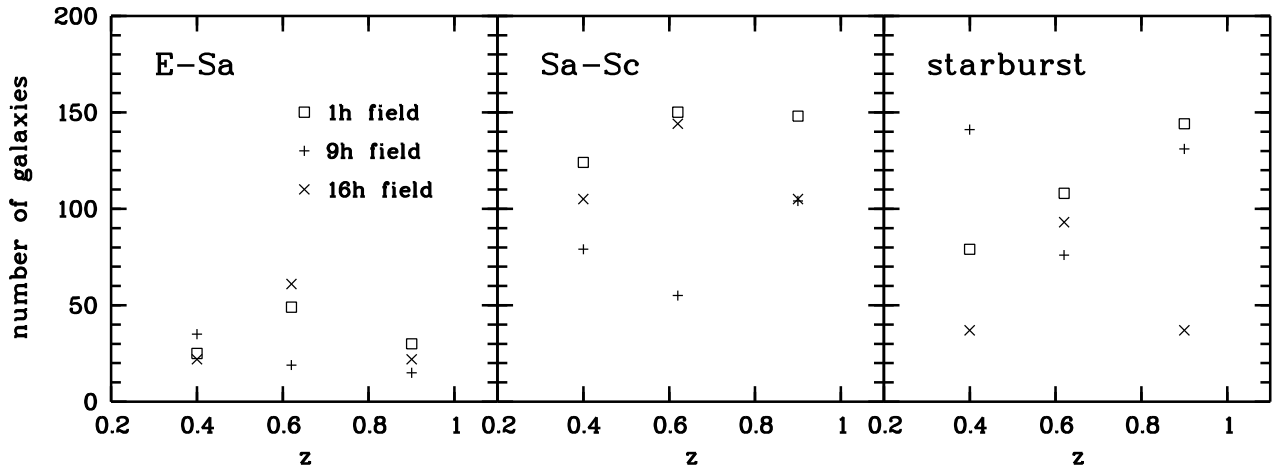}}

    \caption[]{The field to field variations of the number of galaxies used in the 
determination of the luminosity functions.}
      \label{var}
\end{figure*}

As customary, we have calculated the errors based on Poisson statistics. 
This is correct as long as the total volume sampled by the survey is 
considered, where large-scale structure averages out. The scatter in 
fig.\ref{numcount} agrees with Poisson statistic. However, if the survey is 
devided into smaller subvolumes, cosmic variance due 
to clustering of galaxies increases the actual uncertainties. Fig.\ref{var} shows 
the field to field variations using the same  SED and redshift bins as 
applied in the determination of the luminosity function. The scatter in 
figs.\ref{densevol} and \ref{lumevol} is higher than Poisson by factors 
of a few. If this is taken into account, the decrease  in 
comoving galaxy space density for the E-Sa galaxies with increasing redshift 
is statistically not significant, the increasing space density of
the Sa-starburst galaxies towards earlier epochs, however, still is significant. We emphasize 
that this problem of clustering exists with all previous surveys; 
quite contrary, our survey is much larger and will be still larger 
once all targeted fields are available.

\begin{table*} \label{tab_lumfun}
\caption [ ]{Parameters of Schechter function fits to the data.} 
\begin{flushleft}
\begin{tabular}{lllll} 
\hline 
     &    & $\Omega_m=1$   &       \\ 
      
sample         &   z range             &           $\phi*$                                                            & $M^*$                           & $\alpha$  \\ \hline         
E-Sa &0.3-0.5 & $ 0.901 \, 10^{-2} \pm 0.559 \, 10^{-3}$&$-19.287 \pm 0.219$&$0.187\pm 0.224 $  \\ 
E-Sa &0.5-0.75 &$ 0.746 \, 10^{-2} \pm 0.487 \, 10^{-3}$&$-19.355 \pm 0.180$&$-0.25\, 10^{-2} \pm 0.194 $ \\ 
E-Sa &0.75-1.04&$ 0.455 \, 10^{-2} \pm 0.562 \, 10^{-3}$&$-19.360 \pm 0.337$&$0.282\pm 0.52 $  \\ 

Sa-Sc &0.3-0.5&$ 0.137 \, 10^{-1} \pm 0.355 \, 10^{-2}$&$-19.660 \pm 0.295$&$-0.815\pm 0.129 $  \\ 
Sa-Sc &0.5-0.75&$ 0.159 \, 10^{-1} \pm 0.216 \, 10^{-2}$&$-19.335 \pm 0.158$&$-0.615\pm 0.128 $  \\ 
Sa-Sc &0.75-1.04&$ 0.182 \, 10^{-1} \pm 0.462 \, 10^{-2}$&$-19.770 \pm 0.268$&$-0.807\pm 0.295 $  \\

starburst &0.3-0.5&$ 0.177 \, 10^{-1} \pm 0.715 \, 10^{-2}$&$-18.690 \pm 0.292$&$-1.282\pm 0.209 $  \\ 
starburst &0.5-0.75&$ 0.281 \, 10^{-2} \pm 0.273 \, 10^{-2}$&$-20.27 \pm 0.400$&$-1.857\pm 0.168 $  \\starburst &0.75-1.04&$ 0.265 \, 10^{-1} \pm 0.102 \, 10^{-1}$&$-19.441 \pm 0.227$&$-1.727\pm 0.271 $ \\ \hline

  & &$\Omega_m=0.3, \Omega_{\Lambda}=0.7$ & & \\ 
sample         &   z range             &           $\phi*$                                                            & $M^*$                           & $\alpha$  \\ \hline         

E-Sa &0.3-0.5 & $ 0.488 \, 10^{-2} \pm 0.275 \, 10^{-3}$&$-19.602 \pm 0.178$&$0.207\pm 0.178 $  \\ 
E-Sa &0.5-0.75 &$ 0.315 \, 10^{-1} \pm 0.332 \, 10^{-3}$&$-19.871 \pm 0.276$&$-0.475\, 10^{-1} \pm 0.221 $ \\ 
E-Sa &0.75-1.04&$ 0.144 \, 10^{-2} \pm 0.301 \, 10^{-3}$&$-19.705 \pm 0.316$&$0.625\pm 0.587 $  \\ 

Sa-Sc &0.3-0.5&$ 0.831 \, 10^{-2} \pm 0.139 \, 10^{-2}$&$-19.911 \pm 0.193$&$-0.782\pm 0.892 \, 10^{-1} $  \\ 
Sa-Sc &0.5-0.75&$ 0.697\, 10^{-2} \pm 0.971 \, 10^{-3}$&$-19.832 \pm 0.159$&$-0.605\pm 0.159 $  \\ 
Sa-Sc &0.75-1.04&$ 0.676 \, 10^{-2} \pm 0.115 \, 10^{-2}$&$-20.232 \pm 0.194$&$-0.667\pm 0.219 $  \\

starburst &0.3-0.5&$ 0.119 \, 10^{-1} \pm 0.360 \, 10^{-2}$&$-18.970 \pm 0.233$&$-1.141\pm 0.184 $  \\ 
starburst &0.5-0.75&$ 0.211 \, 10^{-2} \pm 0.162 \, 10^{-2}$&$-20.365 \pm 0.299$&$-1.730\pm 0.147 $  \\starburst &0.75-1.04&$ 0.108 \, 10^{-1} \pm 0.329 \, 10^{-2}$&$-19.892 \pm 0.203$&$-1.480\pm 0.276 $ \\ \hline

\end{tabular}
\end{flushleft}
\end{table*}

\begin{table*} \label{evol}
\caption [ ]{Parameters of linear fits of the form $a+b(1+z)$ to the comoving galaxy density 
and B-band luminosity density. Absolute terms $a$ are given in the 2., slopes $b$  in the 
3. column } 
\begin{flushleft}
\begin{tabular}{lll} 
\hline 
        & $\Omega_m=1, \Omega_{\Lambda}=0, \Omega_{\Lambda}=0$   &       \\ 
              &         a       & b \\ \hline
number density E-Sa  &$ 0.669 \, 10^{-2} \pm 0.192 \, 10^{-2}  $&$ -0.174 \, 10^{-2} \pm 0.111 \, 10^{-2} $\\ 
number density Sa-Sc &$ 0.681 \, 10^{-3} \pm 0.339 \, 10^{-2} $&$ 0.744 \, 10^{-2} \pm 0.205 \, 10^{-2} $\\
number density starburst &$ -0.216 \, 10^{-1} \pm 0.191 \, 10^{-2} $&$ 0.184 \, 10^{-1} \pm 0.129 \, 10^{-2} $\\
luminosity density E-Sa &$ -0.138 \, 10^8 \pm 0.163 \, 10^8  $&$ 0.354 \, 10^8 \pm 0.103 \, 10^8 $\\
luminosity density Sa-Sc&$ -0.153 \, 10^9 \pm 0.300 \, 10^8  $&$ 0.179 \, 10^9 \pm 0.193 \, 10^8 $\\ 
luminosity density starburst&$ -0.198 \, 10^9 \pm 0.163 \, 10^8  $&$ 0.167 \, 10^9 \pm 0.113 \, 10^8 $\\ \hline 

    &$\Omega_m=0.3, \Omega_{\Lambda}=0.7$ & \\ 

            &         a       & b \\ \hline

number density E-Sa  &$ 0.630\, 10^{-2} \pm 0.104 \, 10^{-2}  $&$ -0.289 \, 10^{-2} \pm 0.583 \, 10^{-3} $\\ 
number density Sa-Sc &$ 0.871 \, 10^{-2} \pm 0.181 \, 10^{-2} $&$ -0.175 \, 10^{-2} \pm 0.105 \, 10^{-2} $\\
number density starburst &$ 0.630 \, 10^{-2} \pm 0.104 \, 10^{-2} $&$ -0.269 \, 10^{-2} \pm 0.583 \, 10^{-3} $\\
luminosity density E-Sa &$ 0.309 \, 10^8 \pm 0.100 \, 10^8  $&$ -0.640 \, 10^7 \pm 0.582 \, 10^7 $\\
luminosity density Sa-Sc&$ 0.899 \, 10^7 \pm 0.164 \, 10^7  $&$ 0.317 \, 10^8 \pm 0.989 \, 10^7 $\\  
luminosity density starburst&$ -0.756 \, 10^8 \pm 0.878 \, 10^7  $&$ 0.659 \, 10^8 \pm 0.567 \, 10^7 $\\ \hline

\end{tabular}
\end{flushleft}
\end{table*}

\subsection{Comparison with other work}

Since the luminosity function depends strongly on galaxy type, 
a comparison between results obtained from samples selected with 
different selection criteria is difficult if not impossible. 
In a similar way, division into differing subsamples causes differences 
in the luminosity function. Furthermore, different ways to 
estimate K-corrections can introduce redshift dependent effects. 
Photometric errors should be negligible if CCD data are used.

The CFRS ( \cite{lilly95})  is the survey which compares  
most directly to CADIS: their objects are also I-band selected 
across an identical redshift range. Since the number of galaxies 
included in our analysis is nearly four times as large, we have better statistics. 
Furthermore, our survey goes deeper by $0.5 mag$ and so 
the range in luminosities covered is larger, which enables better fits
of  the Schechter function. When we divide our sample 
in the same two SED bins as CFRS, our findings agree very well with their main 
conclusions, namely no evolution of the early type galaxies (E-Sbc) 
and brightening and steepening of the luminosity function with redshift 
for later types.   Inspection of the data points  shows that 
the two surveys agree within the errors (though our errors are smaller).

\cite{lilly96} have determined the evolution of the luminosity 
density from the CFRS data. For the $z=0.35$ bin they obtained 
$log L = 19.48 W Hz^{-1} $ (the value we use here is their 
'directly observed' 4400 value, converted to $H_0 = 100$). 
Using $L_{\odot}(B) = 3.4 \, 10^{11}\, W Hz^{-1}$ we obtain   
$log\, L = 19.38 \, W Hz^{-1} $ at $z=0.35$ from the fits in table \ref{evol}.  
This value includes only galaxies brighter than $M_B=-18.5$; the correction 
factor to the full population down to $M_B=-14$ determined in the two lowest 
redshift bins is a factor of 2, resulting in $log L = 19.68 W Hz^{-1} $ 
which agrees within the errors with the value obtained by \cite{lilly96}. 
At $z=1$  \cite{lilly96} find an increase in comoving luminosity density 
of the blue galaxies by a factor of 2.9 over $z=0.35$ compared to our data 
which give a factor of 3.4 for the late types, again well within the errors. 
We therefore conclude that our deeper and larger redshift survey is  in 
excellent agreement with the CFRS.

The CNOC2 \cite{cnoc2} sample is R-band selected and complete 
to $R=21.5$. It contains about 2000 galaxies with redshifts 
$0.12 z < 0.55$. Their luminosity functions do not agree well with 
those of the CFRS (see \cite{cnoc2}) and hence also not with ours.  

Possible explanations for differences to CFRS have been discussed by 
\cite{cnoc2}, but the cause remains unclear. It seems 
plausible that sample selection effects and k-corrections 
do at least partly  cause the differences, since the CNOC2 
sample is R-band selected whereas CFRS and CADIS are I-band selected. 

The interpretation of the CNOC2  data is based on the evolution of the 
parameters of the Schechter function; since these are coupled, the 
parameters describing the evolution are coupled, too, which may give misleading results 
(see below). Also, \cite{cnoc2} keep $\alpha$ fixed, although the luminosity functions 
of the late type (starburst) galaxies clearly show steepening. One should 
further note that the CNOC2 results are obtained over a smaller redshift 
range than CFRS or CADIS. Additionally, the errors of the luminosity function 
determinations in all available surveys are large and the significance of the results is low.

The autofib survey (\cite{ell96}) is a blue selected sample 
and contains redshifts for $\approx 1700$ galaxies. A steepening of the 
luminosity function from redshifts $z=0.02$ to $z=0.75$ has 
been claimed, although the error ellipses for the lowest and highest 
redshifts bins (Fig. 11 in \cite{ell96}) overlap on the $3\sigma$ level.

\section{Discussion}

We have measured the evolution of the luminosity function; its 
traditional interpretation in terms of density vs. luminosity evolution 
is problematic in the context of a hierarchical picture. 
For the early type galaxies E-Sa,  both $\phi^*$  and the comoving space density 
appear to decrease with redshift, albeit statistically significant only if clustering of 
galaxies is neglected. Since $M^*$ and $\alpha$ are constant, this result implies 
that there are fewer early type galaxies at higher 
redshift. This is exactly what one would expect in hierarchical clustering scenarios 
which require that massive objects form late by  merging of smaller objects. 
Opposed to this scenario is the early formation of galaxies at high redshift 
in a collapse and a single burst of star formation followed by passive evolution. 
Observational evidence  has been found for either scenario 
(see \cite{scha99} for a detailed discussion). From our data, the  E-Sa galaxy 
space density at  $z=1$ is lower by a factor of 1.6 ($q_0=0.5$) compared to $z=0$. 
This is close to the factor of 2-3   predicted by  hierarchical clustering models 
(\cite{kauff96}, \cite{bau96}). The increase of the space density of the 
Sa-Sc and starburst galaxies with redshift fits into this picture as well.

Though our data clearly show evolution of the galaxy population, the 
interpretation of the data is not straightforward. The luminosity function 
describes the galaxy population in a statistical way and contains no direct 
information on the evolution of individual galaxies.  A galaxy may 
disappear from  a given class (or SED bin) either because it disappears 
(e.g. through merging) or  because it changes colors (e.g. by starburst) and 
therefore appears in another bin.  Therefore data like the ones presented here 
{\em alone} do not allow to distinguish between luminosity and density evolution.  
Furthermore, setting a magnitude limit in one pass-band may introduce 
systematic redshift dependent effects. Hence, modelling of different 
evolutionary scenarios  including the detection and classification 
process of the survey is required. In a forthcoming paper we will 
compare our results with hierarchical semianalytical models of the 
formation and evolution of galaxies.

For all disk galaxies (Sa-Sc and starburst) our data show clear evolution. 
It is already well established that disk galaxies evolve to the present day. 
Late type disk galaxies have present-day star formation rates comparable 
to those at earlier cosmic times, early type disks have formed most 
stars at earlier times (\cite{kenn94}). Brightening of galactic disks by 
$\approx 2 mag$ at $z=1$ has been found  by \cite{scha96} which they 
interpret as luminosity evolution. \cite{mao1998} have  additionally used 
kinematic data and interpreted the disk brightening as decrease in 
disk scale length $\simeq (1+z)^{-1}$.  We note that the purely photometric 
classification of objects used in CADIS is less susceptible to 
systematic redshift dependent effects than surveys which use morphological 
criteria. However, only detailed modelling of the detection process can show 
how selection effects affect the data.

HST imaging data are now available for a fair number of galaxies 
drawn from the CFRS and LDSS samples with known redshifts 
(\cite{lil98}, \cite{bri99}, \cite{scha99}, \cite{lef2000}). 
A direct comparison with these studies is difficult because 
there is only a loose  correlation between our spectroscopic 
and their morphological  classification. However, almost certainly a major 
fraction of our starbursting galaxies  would be 
morphologically classified as  irregular and/or  merging. 
\cite{bri99} have found an increase in the proportion 
of irregular galaxies  and \cite{lef2000} found an increase in the merger 
fraction with redshift.  So the evolution of the starbursting galaxy population 
with redshift  is supported by the imaging results.

\section{Conclusions}

We have determined the luminosity function of 2779 field galaxies based on 
multi-color data from the CADIS survey. The data base contains 9-12 color 
indices from optical to IR wavelengths for all objects from which a spectroscopic 
classification according to the SED and a redshift are derived by comparison with 
libraries for stellar, galaxy and QSO spectra. Our data are complete to 
$I_{815}=23.0$ and extend to redshifts $z=1.04$. Our main findings are:

1. The evolution of the B-band luminosity function is clearly 
differential. We find the normalization $\phi^*$ of the early type E-Sa 
galaxy luminosity function and the integrated comoving space density 
to be decreasing with increasing redshift although the latter with less significance. 
The normalization for the Sa-Sc galaxy luminosity function 
increases with redshift as well as the comoving space density. 
The luminosity function of the starburst galaxies steepens towards the faint end 
and their comoving space density increases with redshift.

2. The  survey most directly comparable with our data base is the CFRS. 
When we divide or sample in the same galaxy types, the results 
agree very well. The two other samples which have studied the evolution 
of the luminosity function are not directly comparable to our data; 
we confirm the steepening of the luminosity function found in the autofib 
survey, but not all evolutionary effects claimed by the CNOC2 survey.

3. The density evolution of the early and late type galaxy population 
apparent in our data is suggestive of merging. However, since merging 
increases the star formation rate and consequently the B-band 
luminosity, the interpretation  is not straightforward and requires 
comparison to models.

\begin{acknowledgements}
We thank  S.V.W.Beckwith for his continued support of CADIS. We  would 
also like to thank M.Alises and A.Aguirre for carefully  carrying out 
observations in service mode.
\end{acknowledgements}

\clearpage

\end{document}